# CONTROL OVER MULTISCALE SYSTEMS WITH CONSTRAINTS
# 3. GEOMETRODYNAMICS OF THE EVOLUTION OF SYSTEMS WITH VARYING CONSTRAINTS

**S. Adamenko, V. Bolotov, V. Novikov**

With the use of the general variational principle of self-organization of systems with varying constraints, namely *the principle of dynamical harmonization of systems* presented in the first work of the cycle, we advance an approach to the control over the evolution of systems of many particles. The geometric nature of this principle is analyzed. On the basis of the de Broglie--Bohm representation of the Schrödinger equation, we establish a connection of the nonlocality and the coherence of the systems of many particles with mass entropic forces. The defining role of a coherent acceleration and a space-time curvature in the control over the synthesis of new structures in systems with varying constraints is demonstrated. The basic criteria for electromagnetic fields to initiate the processes of self-organizing synthesis and for the quantum properties of a nonlocality on macroscopic scales, which are necessary for the self-organizing synthesis, are formulated.

**CONTENTS**

## 1. INTRODUCTION

This work is a sequel of the cycle of works (see [1-2]), where some approach to the control over the evolution of the systems of many particles on the basis of the general variational principle of self-organization (*the principle of dynamical harmonization of systems*) is presented. The purpose of the cycle is the development of foundations of the theory and the technology of the synthesis of final states of a system of particles with desired structure and energy binding, which are attained from a given initial state with the help of the initiation of a natural evolution and the control over an evolutionary trajectory of the system at the expense of its internal power resources at a minimal use of the energy of external drivers.

The purpose of the present work is the determination of criteria of the initiation of the self-organizing synthesis, classification of needed drivers, and development of the теории of control over the processes of synthesis on the basis of using the geometric nature of the evolution in the frame of the variational principle of dynamical harmonization.

As is known, the variational principles are the most general and brief means to formulate the laws of the Nature. For example, the equations of dynamics of a system of particles follow under very general conditions from the Gauss least-compulsion principle [3], and the equations of Maxwell and Einstein can be derived from the principle of least action [4].

Our purpose requires us to solve a strongly nonlinear optimization problem. In this problem, it is necessary, in fact, to optimize the trajectory of a system and to appropriately modify the conditions of optimization of this trajectory. Here, we will substantite a possibility of the power-informational control over the evolution of an ensemble of many particles in a noninertial reference system.

In the case under consideration, the essential point is a control nonlinearity related to the fact that the evolution of a system of particles (changes of its structure and the energy of constraints) and the space-time metric are mutually dependent. The theory of evolution of the systems with regard for their internal and external geometries, which will be developed in the present work, can be called the geometrodynamics of the evolution of the systems with varying constraints.

The principle of dynamical harmonization [1, 5] asserts that the self-organization of a system of particles, being under the action of mass forces leading to coherent accelerations of all particles of the system, is directed always to the realization of the transition from the initial state to a state with maximally free dynamics by means of changes of the structure of the system and its inertia relative to mass forces.

According to the Gauss least-compulsion principle, we should vary the accelerations of particles at a fixed velocity lying in a plane tangent to the trajectory. Hertz noticed that the varied accelerations can be related to inertial forces (mass forces) and showed for some simple cases that the minimum of the Gauss compulsion function is equivalent to the minimum of the curvature of the trajectory of a particle. This look at one of the most general variational principles of mechanics allowed one to develop the geometric interpretation for it: the trajectories of particles are geodesic lines in a space.

The idea of the geometrization of the laws of physics was intensively developed in the last century. The most remarkable example of the geometrization of physics is Einstein's general relativity theory, which established the continuous connection of geometry and matter. The fields of gravitation (it is a mass force) induce coherent accelerations and form a curvature of the space-time, where the particles are moving freely and are simultaneously the sources of the curvature of this space-time. In other words [6], "matter tells space how to curve, and space tells matter how to move." It is of importance that such an approach to the field theory allowed one not only to describe the fields of gravitation, but also to deduce the equations of motion of particles directly from the field equations [7,8], if the idea of particles as the singular solutions of the field equations is used.

The idea of particles related to singularities was somewhat earlier introduced by L. de Broglie, who tried to interpret the quantum-mechanical dynamics of particles in the frame of his theory of double solution [10-11] on the basis of the Madelung hydrodynamic representation [9] for the Schrödinger equation. In this quantum-mechanical theory, it is proposed to represent the dynamics of a particle by the sum of two solutions of волнового the Schrödinger wave equation, namely the smooth and singular ones.

The importance of the notion of nonlocality for the theory of self-organization was indicated in the first work of the cycle [1]. Here, we will refine the connection of the property of nonlocality of the wave functions determined from the Schrödinger equation (see [12]) with mass and entropic forces and will show that this property is also revealed in classical physics as a result of the geometrization of the physical processes of dynamics and evolution.

In the frame of classical physics, the geometrization of the dynamics of particles, which is garmonically associated with the property of nonlocality, was first realized by A. Vlasov. He constructed a nonlocal statistical theory [13-15] and obtained kinetic equations on the basis of the geometry of a space of support elements. "The space of support elements" includes the following notions:
1. Coordinate space.
2. Tangent space and the tangency order.
3. System of vectors loaning on the tangency point and lying in the tangent space.

The space of support elements joins the coordinate space (as the space of the possible values of the centers of mass of particles) and as the space of the possible values of kinematic parameters of particles, for example, their velocities, ans also can include the vectors of accelerations of an arbitrarily large order, which depends on the tangency order.

The validity of the space of support elements consists in the exact formation of a new understanding of a particle, which is characterized by the continuum of the possible values of coordinates and velocities (and also accelerations of any order), as distinct from the classical image of a localized particle with definite values of coordinates and velocities [14].

## 2. SCHRÖDINGER EQUATION AND ENTROPIC FORCES

The most efficient apparatus for the analysis of states of the system undergoing coherent accelerations is presented by the Schrödinger equation and covariant kinetic equations.

The Schrödinger equation for the wave function $\Psi(\vec{r}_1, \vec{r}_2, ..., \vec{r}_N, t)$ describing the ensemble of $N$ particles,

$$i\hbar \frac{\partial \Psi(\vec{r}_1, \vec{r}_2, ..., \vec{r}_N, t)}{\partial t} = -\frac{\hbar^2}{2m} \sum_{n=1}^{N} \nabla_n^2 \Psi(\vec{r}_1, \vec{r}_2, ..., \vec{r}_N, t) + U(\vec{r}_1, \vec{r}_2, ..., \vec{r}_N) \Psi(\vec{r}_1, \vec{r}_2, ..., \vec{r}_N, t), \quad (1.1)$$

where $\nabla_n = \left\{ \frac{\partial}{\partial x_n} \vec{i} + \frac{\partial}{\partial y_n} \vec{j} + \frac{\partial}{\partial z_n} \vec{k} \right\}$, and $U(\vec{r}_1, \vec{r}_2, ..., \vec{r}_N)$ is the potential of external forces acting on particles, arose from the attempt to solve some problems of the dynamics of particles on small spatial scales. The linear equation for complex-valued wave functions, which was obtained as a generalization of the classical dynamics of particles characterized by real variables, has shown a very good agreement with experiment. The transition from complex-valued variables in the Schrödinger equation conversely to real variables in the frame of the de Broglie--Bohm representation (see [9, 11, 16]),

$$\Psi(\vec{r}_1, \vec{r}_2, ..., \vec{r}_N, t) = \sqrt{\rho(\vec{r}_1, \vec{r}_2, ..., \vec{r}_N, t)} \cdot \exp\left(i \frac{J(\vec{r}_1, \vec{r}_2, ..., \vec{r}_N, t)}{\hbar}\right), \quad (1.2)$$

where $\rho$ - probability density, and $J$ - action, allowed one to show that the differences between classical and quantum mechanics are reduced to the appearance of an additional potential $U_q(\vec{r}_1, \vec{r}_2, ..., \vec{r}_N)$ in the system of equations for real variables. It was called the "quantum" potential (see the details in Section 2).

In the present work on the basis of the representation of de Broglie and Bohm for the wave function of a system of particles with regard for the potential $U_q$, we will obtain the equations for the entropy and the formulas for coherent accelerations in electromagnetic fields. We will analyze the solutions of kinetic equations with regard for quantum statistics for an ensemble of particles undergoing the coherent acceleration and obtain the solutions of kinetic covariant equations in noninertial reference systems.

Recently, the more and more attention was paid to the macroscopic objects possessing some quantum properties (in particular, the properties of coherence and nonlocality). In 2001, the Novel Prize in physics was awarded for the creation and study of the Bose-condensates of atomic complexes [17]. In view of the importance of the notion of a "macroscopic quantum object" (MQO), which is considered as a set of particles forming a collective system of macroscopic sizes and possesses the property of nonlocality typical of quantum objects.

The basic property of the evolution turns out to be the space-time anisotropy related to the fact that the factor defining the evolution is an acceleration, and the properties of the space in the directions along the acceleration and perpendicularly to it are obviously different. Such an anisotropy corresponds to the obtained solutions of covariant equations that have power asymptotics and define the fast localization of the domain of existence of the system in the direction of action of a mass force and its fast delocalization in a subspace orthogonal to the direction of the acceleration.

In this case, the natural geometry of the space-time is the geometry of the space of support elements (the support elements are the kinematic elements on the trajectory of a particle), namely the Finsler geometry.

Such a viewpoint combines, in fact, the approaches of L. de Broglie, J.-P. Vigier, A. Einstein, and A. Vlasov concerning the geometric nonlinear nature of the physical laws of dynamics and evolution. The role of a support is played by the four-dimensional space-time. The tangent bundles are the planes of accelerations of all orders and the space of the internal structure of a system, i.e., the space with coordinates characterizing a structure of constraints in the system (e.g., such as the fractal dimension or the order parameter, deformations, etc.). The tangent bundles and the support, which is a space-time with Riemann geometry, are coupled by the vector of acceleration or the space-time curvature.

As for the relationship of a self-organizing system and the space-time, we note that
- the systems of particles, by undergoing the coherent accelerations during the evolution, make the space-time, where they are placed, curved;
- in turn, the space-time becoming curved indicates the directions of free motion for particles and the directions of evolution of the internal structure of the system.

The obtained solutions of covariant kinetic equations under a fast extension of the space-time becoming curved are similat to the accelerated cosmological extension in the general relativity theory (see [18]) with cosmological constant. As became clear in the 1960s [19-21], such an extension is continuously connected with the physical vacuum, which has the antigravity properties corresponding to the cosmological term in the Einstein equation.

The physical vacuum has many significant properties. In particular, it was established that vacuum is homogeneous on scales from centimeters or meters up to cosmic scales. On the "ordinary" and subatomic scales, the homogeneity of vacuum can be broken with the appearance of experimentally observed macroscopic effects related to the polarization of vacuum (Casimir effect) and to the coherent acceleration (dynamical Casimir effect) [22-24]. Near (and inside) the systems that undergo phase transitions and coherent accelerations, the space-time becomes curved (see the Vlasov theory [13-14]), and the light velocity is changed (see experimental results in [25]).

Here, we consider the possibilities to use the electromagnetic fields and the fields of entropy gradients for the control over the evolution of a system, i.e., over the evolution of its constraints. In this case, the space-time curvature arises, and, hence, the resonances of mass forces в создаваемых ими неоднородностях of the space-time and the physical vacuum can appear. These resonances are analyzed, and their parameters are determined.

The application of these resonances to the control over the synthesis of systems with varying constraints can become a promising element of the future technologies with "guided evolution".

The natural consequence of the space-time curvature is the difference of the intrinsic time from the laboratory one. The former depends not on the velocity of the reference system, but on its acceleration, which leads to a change in the lifetime of particles [26].

Similar effects were noticed by Vlasov [14] and Kozyrev [27], and the influence of the growth of crystals on the light velocity was experimentally discovered as early as 1905 [25].

It is especially interesting that Kozyrev was able, starting from his theory of time, to fabricate a special gage on the basis of resistors forming a bridge scheme and to observe the motion of stars at a laboratory on the Earth in real time (see, e.g., [28]). In our opinion, these effects are related to the appearance of a local space-time curvature determining the ratio of the intrinsic and laboratory times, as well as the impedance of electrotechnical elements (see, e.g., [29]).

The close ideas were developed in works by S. Podosenov (see [29]), where he showed how the constraints in a system determine the Riemann space-time curvature.

In the present work, we will show that the variational principle of dynamical harmonization leads to the geometrization of the physical processes of evolution and to a generalization of the theory for any fields of mass forces. We give the experimental results obtained with the use of the Kozyrev detector. We consider that they demonstrate, under laboratory conditions, the resonances of electromagnetic radiation with the inhomogeneity related to the space-time curvature initiated by the action of electromagnetic drivers described below.

The resonances caused by the propagation of longitudinal waves amplify the fluctuations of vacuum under conditions of the scale invariance and, acting on particles, initiate their complicated motion, which can develop into a dynamical chaos. As was shown in [30], it is convenient to apply Tikhonov's methods of regularization of the states with dynamical chaos to the description of the dynamics of particles. The scale invariance of vacuum implies that the operators of regularization can be given in terms of quantum [31] and fractional [32] integro-differential operators. The developed model of the interaction with vacuum describes naturally the openness of the system and the processes of creation and decay [33].

The equations of the dynamics of particles at their interaction with the scale-invariant vacuum after the regularization can be modeled with electrophysical curcuits: operational amplifiers as a model of the operators of regularization and branched equivalent resonance schemes as a model of fractal medium. The developed model is especially useful for the optimization of electromagnetic drivers.

We emphasize once more that a specific feature of the considered technology of the control over the evolution of systems is the use of the internal mass-defect energy in order to change a structure of the system, rather than the energy of external drivers. The low energy of external fields must be spent only for the control and the initiation of the processes of self-organization with desired structural and energetic directedness. In what follows, we will show that namely the nonlocal entropic fields determine the main properties of MQO.

**2.1 The de Broglie--Bohm representation for the Schrödinger equation and entropic forces**

The main property of MQO (i.e., a quantum system co constraints) consists in that the wave function describing it cannot be represented as a product of one-particle wave functions in the form

$$\Psi(\vec{r}_1, \vec{r}_2, ..., \vec{r}_N, t) = \Psi(\vec{r}_1, t) \Psi(\vec{r}_2, t) \cdots \Psi(\vec{r}_N, t). \tag{2.1}$$

For this function, the normalizing integral is reduced to product of independent integrals:

$$\int_{V_1} \Psi^*(\vec{r}_1, t) \Psi(\vec{r}_1, t) dr_1 \cdot \int_{V_2} \Psi^*(\vec{r}_2, t) \Psi(\vec{r}_2, t) dr_2 \cdots \int_{V_N} \Psi^*(\vec{r}_N, t) \Psi(\vec{r}_N, t) dr_1 = 1. \tag{2.2}$$

Each integral in this expression is separately equal to 1. In other words, the behavior of each particle в the ensemble is described by the own wave function independent of the states with other wave functions. Hence, the coupling of particles in the single ensemble is ensured only due to the potential of external fields $U(\vec{r}_1, \vec{r}_2, ..., \vec{r}_N)$ and to the quantum potential $U_q(\vec{r}_1, \vec{r}_2, ..., \vec{r}_N)$ [34]. This coupling is supported by quanta ensuring the given interaction. The rate of exchange by quanta does not exceed the light velocity, as distinct from the entangled states representing MQO, where the information propagates, as will be shown below, instantly.

This can be easily seen from the formula for the probability to find MQO in a given volume $V$:

$$\int_V \Psi(\vec{r}_1, \vec{r}_2, ..., \vec{r}_N, t)^* \Psi(\vec{r}_1, \vec{r}_2, ..., \vec{r}_N, t) dr_1 dr_2 \cdots dr_N = 1. \tag{2.3}$$

In this case, the wave function describing MQO depends on the time, but the probability to find MQO in the given volume is conserved. This implies that any change in the position of any of $N$ particles placed in a volume $V$ affects instantly the positions of all remaining bound particles.

Thus, the instant rearrangement of the positions of all particles reflects the presence of strong constraints in MQO and, hence, strong correlations in it. MQO is a macroscopic formation (most probably, a quasicrystal), which is described by the quantum Schrödinger equation.

A bound dynamical system tends to instantly become self-organized and to pass in a state with maximal probability. For the viewpoint of the principle of dynamical harmonization, the system chooses optimally the direction of a change of the structure formed by entropic fields. This reasoning can be easily generalized to the case of a partially bound dynamical quantum system.

With regard for the de Broglie--Bohm representation (1.2), Eq. (1.1) yields

$$-\frac{\partial J}{\partial t} \Psi + i\hbar \frac{1}{2\rho} \frac{\partial \rho}{\partial t} \Psi = \frac{1}{2m} \sum_{n=1}^{N} (\nabla_n J)^2 \Psi + U(\vec{r}_1, \vec{r}_2, ..., \vec{r}_N) \Psi -$$

$$-\frac{i\hbar}{2m} \sum_{n=1}^{N} \nabla_n^2 J \cdot \Psi - \frac{i\hbar}{2m} \sum_{n=1}^{N} \left(\frac{\nabla_n \rho}{\rho}\right)(\nabla_n J) \cdot \Psi - \frac{\hbar^2}{2m} \sum_{n=1}^{N} \left(\frac{\nabla_n^2 \rho}{2\rho}\right) \Psi + \frac{\hbar^2}{2m} \sum_{n=1}^{N} \left(\frac{\nabla_n \rho}{2\rho}\right)^2 \Psi. \tag{2.4}$$

We note that the probability density and the action are real functions. We can separately collect the real and imaginary terms and obtain two nonlinear equations corresponding to one linear Schrödinger equation:

$$-\frac{\partial J}{\partial t} = \frac{1}{2m} \sum_{n=1}^{N} (\nabla_n J)^2 + U(\vec{r}_1, \vec{r}_2, ..., \vec{r}_N) + \frac{\hbar^2}{2m} \sum_{n=1}^{N} \left(\left(\frac{\nabla_n \rho}{2\rho}\right)^2 - \frac{\nabla_n^2 \rho}{2\rho}\right), \tag{2.5}$$

$$-\frac{\partial \rho}{\partial t} = \sum_{n=1}^{N} \nabla_n \cdot \left( \frac{\rho \nabla_n J}{m} \right). \tag{2.6}$$

We now introduce the entropy density of a quantum system in the form

$$S(\vec{r}_1, \vec{r}_2, ..., \vec{r}_N, t) = -\ln \left| \Psi(\vec{r}_1, \vec{r}_2, ..., \vec{r}_N, t) \right|^2 = -\ln \left( \rho(\vec{r}_1, \vec{r}_2, ..., \vec{r}_N, t) \right). \tag{2.7}$$

Let us transform formulas (2.5) and (2.6), by substituting entropy (2.7) in them. In this case, we take into account that $\rho^{-1} \nabla \rho = \nabla \ln \rho$. We have

$$-\frac{\partial J}{\partial t} = \frac{1}{2m} \sum_{n=1}^{N} (\nabla_n J)^2 - \frac{\hbar^2}{8m} \sum_{n=1}^{N} (\nabla_n S)^2 + U(\vec{r}_1, \vec{r}_2, ..., \vec{r}_N) + \frac{\hbar^2}{4m} \sum_{n=1}^{N} \nabla_n^2 S. \tag{2.8}$$

We note that the modified Hamilton--Jacobi equation (2.8), which passes into the classical Hamilton-Jacobi equation at the standard limiting transition as $\hbar \to 0$, contains the term $\frac{\hbar^2}{8m} \sum_{n=1}^{N} (\nabla_n S)^2 = E_s$. This term is an analog of the entropic-type kinetic energy $E_s$ and is expressed through the gradients of entropic fields. The physical смысл of a kinetic energy of the entropic type consists in the binding of a system of particles, which causes a decrease of their total kinetic energy. The action of the entropic force is directed on the optimization of the system of constraints в dynamical system, by leading it to the most probable state. In other words, the system evolves to a new configurational state with maximal stability. The density gradients of entropic fields are a quantitative characteristic of the probabilistic laws and forces.

In the modified Hamilton--Jacobi equation (2.8), the last term $\frac{\hbar^2}{4m} \sum_{n=1}^{N} \nabla_n^2 S$ corrects the potential energy of the system. The sign of the Laplace operator in this expression reflects a decrease or increase in the potential energy of the bound system on the whole. Thus, the direction of the flows of entropy density gradients determines the final value and the shape of the potential of interaction between particles.

*Thus, it becomes clear that the entropic field is related to the fields of constraints in any quantum system, in particular, in MQO. Moreover, the introduction of entropic forces makes the separation of quantum and classical mechanics, which has born always a sufficiently indefinite character, to be conditional. The border between them is eroded, if we consider the dynamics of classical systems with regard for the evolution of their internal constraints in the presence of the corresponding entropic fields.*

From our viewpoint, it is significant that formula (2.8) contains quantum terms together with classical ones. The former can be expressed in terms of the operators of momentum and kinetic energy of particles of the system: $\hat{p}_n = -i\hbar \nabla_n$, $\hat{T}_n = -\frac{\hbar^2}{2m} \nabla_n^2$.

Using these operators, we can transform Eq. (2.7) to the form

$$-\frac{\partial J}{\partial t} = \frac{1}{2m} \sum_{n=1}^{N} (\nabla_n J)^2 + \frac{1}{8m} \sum_{n=1}^{N} (\hat{p}_n S)^2 + U(\vec{r}_1, \vec{r}_2, ..., \vec{r}_N) - \frac{1}{2} \sum_{n=1}^{N} \hat{T}_n S. \tag{2.9}$$

Formulas of the type (2.10), which include classical terms and the operators of physical quantities, describe the macroscopic quantum objects. Thus, MQOs reveal both classical and quantum properties.

The equation of balance of the entropy follows simply from Eq. (2.6):

$$\frac{\partial S}{\partial t} + \sum_{n=1}^{N} (\vec{u}_n \cdot \nabla_n S) - \frac{1}{m} \sum_{n=1}^{N} \nabla_n^2 J = 0. \tag{2.10}$$

The knowledge of solutions of the system of nonlinear differential equations (2.8), (2.10) for the action and the entropy allows us to write the wave function, being a solution of the Schrödinger equation, in the form

$$\Psi(\vec{r}_1, \vec{r}_2, ..., \vec{r}_N, t) = \exp(-Z), \quad Z = \frac{S(\vec{r}_1, \vec{r}_2, ..., \vec{r}_N, t)}{2k} - i \frac{J(\vec{r}_1, \vec{r}_2, ..., \vec{r}_N, t)}{\hbar}. \tag{2.11}$$

Here, we write the Boltzmann entropy in the form $S = -k_B \ln \rho$, where $k_B$ is the Boltzmann constant. The function $Z$ is a complex-valued function with nonzero real and imaginary parts.

Let us consider the action of the operators of momentum and kinetic energy on the entropy:

$$\hat{p}_n S = \frac{1}{8m} \left( \frac{1}{\Psi^*} (\hat{p}_n \Psi^*) - \frac{1}{\Psi} (\hat{p}_n \Psi) \right). \tag{2.12}$$

If the eigenvalues of the operator of momentum are real numbers, then the action of the operator of momentum on the entropy is equal to zero, but it is possible only in the case where the wave function of the system of particles can be expanded in a product of wave functions.

The action of the operator of kinetic energy on the entropy is given as follows:

$$-\frac{1}{2}\sum_{n=1}^{N}\widehat{T}_n S = \sum_{n=1}^{N} E_n - \sum_{n=1}^{N}\frac{p_n^2}{2m} = E - \frac{P^2}{2m} \qquad (2.13)$$

In this case, the eigenvalues of the operator of kinetic energy are real. We denote such an eigenvalue for the *n*-th particle as $E_n$ and the total momentum of particles of the system as $P$.

It is seen from (2.13) that the action of the operator of kinetic energy on the entropy of the system is since $E = \frac{P^2}{2m}$. This result is possible only under the condition that the wave function of the system of particles can be expanded in a product of one-particle wave functions, which breaks MQO. In this case, the generalized quantum-classical Hamilton-Jacobi equation passes into its classical analog.

The performed calculations allow us to assert that **the quantum corrections related to the entropic fields appear only in the presence of long-range correlations in the systems of particles, i.e., if MQO arises.**

By analogy with classical mechanics, it is easy to determine the momenta that are determined by the mass entropic fields. As is seen from the formula for the entropic-type kinetic energy, each entropic momentum $p_{sn}$ acquired by the $n$-th particle is proportional to the entropy gradient.

$$p_{sn} = \frac{\hbar}{2}\nabla_n S \ . \qquad (2.14)$$

The entropic momentum transfers each of the particles of MQO in the position that corresponds to the maximum of the probability of a state for the given MQO. Thus, the dominating mass force (general dominating perturbation) [5] supplies coherently the momentum $p_{sn}$ of a directed motion to all elements of the ensemble of particles, which meets the condition $\left|\sum_{n=1}^{N}\vec{p}_{sn}\right| \gg P_{T_{\max}}$, where $P_{T_{\max}}$ is the maximal absolute value of the momentum of the intrinsic heat motion of any of the elements of the ensemble (MQO, in this case).

The principle of dynamical harmonization [1] implies that the evolution of a self-organizing system is possible only in the presence of the coherent acceleration of the entire system, when all particles of the system acquire the same momentum increment due to the action of the entropic force arising under the nonzero entropy gradient. In this case, it is necessary that the regular component of the change in the momenta of particles $\Delta p_S$ at the expense of the entropy exceed the chaotic heat component $p_T \approx m u_T$. We call this requirement as the condition of domination of a driver.

It is convenient to introduce the coefficient of domination of a driver $\alpha_d$ as the ratio of the momentum increment of a particle due to the action of a driver to the heat component of the momentum. Then the condition of domination of a driver takes the form

$$\alpha_d \gg 1, \ \alpha_d = \frac{\Delta p}{p_T} \ . \qquad (2.15)$$

The rate of transfer of a momentum to the system of particles allows us to estimate the mass force $F_{str}$ stimulating the system to the coherent acceleration and the evolution due to a change of the internal structure.

Thus, the analysis of the Schrödinger equation implies that the perturbation of MQO related to the appearance of the entropy gradient (mass force) at any point of the volume occupied by MQO causes a change of the momentum of each particle of the given object, since even an insignificant external entropic perturbation acts at once on all particles, which are located at the points $\{\vec{r}_1, \vec{r}_2, ..., \vec{r}_N\}$.

*Such properties of the system indicate the existence of a nonlocality of MQO in the general case. It becomes clear that the physical entropic field (field of mass forces) is the reason for the appearance of the fields of constraints.*

*If condition (2.15) is satisfied, and действии if the system undergoes the action of an entropic dominating perturbation (i.e., if $\nabla S > 2P_{T_{\max}}/\hbar$ is satisfied), the value of "quantum potential" becomes*

***essential for the evolution of the system irrespective of its scale. In other words, under the action of a dominating perturbation, even a classical system becomes nonlocal and acquires quantum properties.***

It is seen from Eq. (2.13) that the entropic fields acting on the system of particles decrease always its kinetic energy. For such systems, we introduce the definition of the degree of nonideality $\Theta$, which shows a share of the decrease of constraints of the system:

$$\Theta = \frac{E - E_s}{E}. \tag{2.16}$$

Here, $E = \frac{1}{2m}\sum_{n=1}^{N}(\nabla_n J)^2 = \frac{1}{2m}\sum_{n=1}^{N} p_n^2$ is the kinetic energy of the system of $N$ particles with the same mass, and $\vec{p}_n = \nabla_n J$.

For $\Theta = 0$, the system is completely bound, has the maximal ideality, and possesses, as a whole, a field of entropy gradients such that leads to a minimum of the kinetic energy. Based on this, we can estimate the mean entropy gradient, due to which the ideality limit and the maximal coherence are attained in the system, i.e., $E \approx E_s$ or $\langle p_s \rangle \approx \langle p \rangle$:

$$\langle \nabla S \rangle \approx \frac{2}{\hbar}\langle p \rangle \tag{2.17}$$

For $\Theta = 1$, the fields of entropy gradients are absent, and the system becomes completely nonideal.

The introduction of the entropic momenta (2.14) leads to a new formula for entropic forces in quantum mechanics,

$$\vec{F}_s = \dot{\vec{p}}_s = \frac{\hbar}{2}\nabla \dot{S}, \tag{2.18}$$

where the dot stands for the differentiation with respect to the time. Hence, the entropic force is proportional to the entropy production gradient $\sigma_S = \dot{S}$ (see also [1]) in the system. This formula differs from that obtained by E. Verlinde [35], who used the equations of equilibrium thermodynamics of closed systems and obtained

$$\vec{F}_s = T\nabla S. \tag{2.19}$$

This formula does not involve the entropy production and, hence, cannot be applied to the description of nonequilibrium systems.

Knowing the entropic force, it is easy to set the entropic pressure into the theory:

$$P_K = \frac{(F_s, \vec{n})}{K}, \tag{2.20}$$

where $K$ — is the area on which the entropic force acts, $\vec{n}$ — is a unit vector of normal to the surface with area $K$. The physical meaning of the entropic pressure is that it forms structures at all hierarchical levels of the system (clusters, molecules, atoms, nuclei, etc.) by changing the space-time structure of the physical vacuum.

In the case of a spherical surface with a radius $R$, the entropic pressure equals:

$$P_K = \frac{\hbar}{2}\frac{\nabla \dot{S}}{4\pi R^2}. \tag{2.21}$$

In order to estimate the entropic pressure in the shell with the thickness $d$ the following formula may apply:

$$\nabla \dot{S} \approx \frac{1}{d}\left(\frac{\Delta S}{\tau}\right), \quad P_K \approx \frac{\hbar}{2}\frac{\Delta S}{4\pi R^2 d}\frac{1}{\tau}, \tag{2.22}$$

where $\tau$ - is the time for the formation of a spherical shell structure at all levels of the hierarchy, $\Delta S$ - is the entropy change in the spherical shell of radius $R$ and thickness $d$.

Taking into account that the entropic pressure has the same value at all levels of the hierarchy, one can determine the change of energy of constrains of the nuclear component of the system. For this to be done it is necessary to equate the work done by the entropic force on the structuring of a spherical shell at nuclear level and the change of the energy of constrains:

$$P_K dV = d\varepsilon, \quad dV = 4\pi R^2 dR \tag{2.23}$$

Thus, the change of the energy of constrains is proportional to the production of entropy:

$$d\varepsilon = \frac{\hbar}{2} d(\sigma_S), \quad \sigma_S = \dot{S}. \tag{2.24}$$

and for the estimation of change of energy of constrains of the system, a relation may be applied:

$$\Delta\varepsilon = \frac{\hbar}{2} \frac{\Delta S}{\tau}. \tag{2.25}$$

This estimate provides a basis to generalize the Heisenberg uncertainty principle for energy and time in systems with varying constrains (due to the change of energy of $\Delta E$):

$$\Delta t \Delta E \approx \frac{\hbar}{2} \Delta S \tag{2.26}$$

Now it is clear that the transition to classical mechanics does not occur while $\hbar \to 0$, which is not actually logical for the constant, but through vanishing of entropy gradient cnange in the system. We now estimate the mass forces and the domination of a driver in the frame of the "shell" theory of evolution (see [1]). Let the action of a driver on the system of particles have lead to the separation of a subsystem of particles (shell) with mass number $A_{sh}$, where the mass force acts.

The coefficient of domination $\alpha_{dS}$ of the action of an entropic driver on a single particle can be represented as

$$\alpha_{dS} = \frac{\hbar(\nabla S / A_{sh})}{m_p} = \left(\frac{\hbar}{m_p}\right) \frac{(\Delta S / A_{sh})}{u_T l_{eff}} \approx 10 \left(\frac{\hbar/(m_p u_T)}{R_{sh}}\right) \Delta S \approx 10 \left(\frac{\lambdabar_{D-B}}{R_{sh}}\right) \frac{\Delta S}{A_{sh}}, \tag{2.27}$$

where $\lambdabar_{D-B}$ is the de Broglie wavelength corresponding to a heat pulse:

$$\lambdabar_{D-B} \approx \hbar / p_T. \tag{2.28}$$

To obtain the final result, we need to estimate a change of the entropy $(\Delta S)_b$ due to the development of physical processes with a change of constraints in the system. On the initial stage, a subsystem of $A_{sh}$ particles is separated. In it, the initial shell structure with the coherent part of $A_{shCog}$ particles and, hence, with the input order parameter $\eta_{sh} = \frac{A_{shCog}}{A_{sh}}$ is created. Then we take into account that

- the break or creation of one constraint consumes the erasure energy of one bit of information $\varepsilon_b \approx T \ln 2$ (Landauer theorem [36]);
- the number of constraints at the formation of a cluster with with mass number $A_{shCog}$ is equal, by the order of magnitude, to $N_{bcog} \approx 0.5(A_{shCog})^2$;
- the probability of the creation of a cluster with the number of particles $A_{shCog}$ is proportional to $P_{cog} \approx 1/\sqrt{A_{shCog}}$;
- $A_{shCog} \approx \eta_{sh} A_{sh}$;
- $T(\Delta S)_b = W_{shCog}$, $V \approx const$, $W_{shCog}$ is the energy of the formation of the coherent part of a shell;
- $W_{sh} \approx P_{cog} N_{bcog} \varepsilon_b$.

The above relations imply that, at the formation of a structure, the entropy is changed by

$$(\Delta S)_b \approx \frac{1}{T} W_{shCog} \approx \frac{1}{T} 0.5 (A_{shCog})^2 \frac{1}{\sqrt{A_{shCog}}} T \ln 2 \approx \frac{\ln 2}{2} (A_{shCog})^{3/2} \approx 0.038 (\eta_{sh} A_{sh})^{3/2}. \tag{2.29}$$

In this case, the specific change of the entropy per particle is

$$\frac{(\Delta S)_b}{A_{sh}} \approx 3.8 10^{-2} \eta_{sh}^{3/2} A_{sh}^{1/2}. \tag{2.30}$$

Substituting this formula in that for the coefficient of domination (2.27), we obtain

$$\alpha_{dS} \approx 0.38 \left( \frac{\lambdabar_{D-B}}{R_{sh}} \right) \eta_{sh}^{3/2} A_{sh}^{1/2} . \tag{2.31}$$

***If the conditions of domination (2.14) are satisfied, the drivers transfer the nucleus of a shell from the quasiequilibrium state in the inertial reference system in a strongly nonequilibrium state in the noninertial system formed under the action of mass forces*** $F_{str} = \dfrac{d\Delta p}{dt}$***. The mass forces cause the coherent acceleration***

$$a_{cog} \approx \frac{1}{m_p} \frac{d}{dt}(\Delta p) \approx \frac{1}{m_p}\left( \frac{\alpha_d p_T}{\tau_{eff}} \right) \approx \alpha_d \frac{u_T}{\tau_{eff}} \tag{2.32}$$

***and the evolution of a structure of the system (see the principle of dynamical harmonization [1]).***

We note that the basic relations have been obtained from the Schrödinger equation in the nonrelativistic case without external electromagnetic field.

**2.2. Electromagnetic drivers of mass forces**

We now show that the electromagnetic fields can be dominating perturbations for a system of particles. As is well known, the Lagrange function of a system of particles in electromagnetic fields is transformed ([37]). Without electromagnetic fields, the momentum of a particle with charge $q$ and mass $m$ is connected with its velocity $\vec{u}$ by the well-known formula $\vec{p} = \dfrac{m\vec{u}}{\sqrt{1-(u/c)^2}}$. At the motion in an electromagnetic field with the vector potential $\vec{A}$ (and in the fields defined by the potential: $\vec{B} = rot\,\vec{A}$, $\vec{E} \propto \dfrac{\partial \vec{A}}{\partial t}$), the total momentum of the particle changes due to the vector potential. Even if there is no magnetic field at the point of the space, where a particle is located, the total momentum of the particle is determined by the formula

$$\vec{P} = \frac{m\vec{u}}{\sqrt{1-(u/c)^2}} + e\vec{A} . \tag{2.33}$$

Similarly to the connection of the electrostatic potential and the energy, the vector potential reveals a connection with the momentum. The vector potential supplies the additional electrodynamic momentum to all charged particles $\Delta \vec{p}_{EM} = e\vec{A}$.

The mass force of the electromagnetic origin, $F_{str} = \dfrac{d(\Delta p)_A}{dt} = q\dfrac{dA}{dt}$, is a force acting on a charge and is given by the derivative of the vector potential with respect to the time in the standard formula for the electric field intensity in terms of the four-dimensional gradient of a four-dimensional potential:

$$\vec{E} = -\nabla \varphi - \frac{\partial \vec{A}}{\partial t} . \tag{2.34}$$

We note that the contribution of the rate of variation of the vector potential
- can essentially exceed that of the electrostatic potential gradient for short pulses;
- can be present in the system even without gradient electric fields and transverse magnetic fields ($rot\vec{A} \approx 0$, and the potential is frequently called a zero-field potential in this case);
- by determining an alternating electric field $\vec{E}$ if the condition $rot\vec{A} \approx 0$ holds, generates no alternating magnetic field, but can ensure the appearance of sources of a vector potential that are concentrated in the regions with $div\vec{A} \neq 0$;
- defines the localization of the magnetic field and sources of a vector potential in spatially remore regions;
- can be present in the system in the case where the electrostatic potential is the same at all points of the system.

There are many means to generate the fields of a vector potential, but such sources as the toroidal coils on a core with magnetic permeability $\mu$ and with current $I_{amper}$ flowing on $n$ windings are most convenient. For such drivers, the amplitude of the vector potential is given by the relation $A \approx \frac{\mu_0}{4\pi} \mu n I_{Amper} = 10^{-7} \mu n I_{Amper}$ (in IS units) and ensures the coefficient of domination

$$\alpha_{dA} = \frac{qA}{m_p u_T} = 10^{-7} \mu \cdot n \cdot \left(\frac{q}{m_p}\right) \frac{I_{Amper}}{u_T}. \qquad (2.35)$$

In correspondence with the equations Maxwell, the component of the electric field intensity $\vec{E} = -\frac{\partial \vec{A}}{\partial t}$ exists also in a homogeneous system of particles, i.e., it can be an electromagnetic mass force acting directly on the charged component of a system of particles.

We now mention one of the simplest drivers. The collective properties of a system of particles are usually revealed, first of all, in the hydrodynamic behavior of the system. It is clear that if the system is affected by a hydraulic impact, then, at its high intensity, the particles receive a momentum increment $\Delta \vec{p}_m$ exceeding the thermal momentum. Moreover, a nonlinear wave moving with supersound velocity appears in the system of particles. It is clear that, in this case, the condition of domination is satisfied on the front of this nonlinear wave is satisfied.

We now summarize the above-performed analysis of the Schrödinger equation: the particles undergo the action of a nonlocal mass force causing a change in the momentum of particles by a value bounded by the sum of the corresponding contributions of basic drivers (mechanical, electromagnetic, and entropic ones):

$$|\Delta p| \leq |\Delta \vec{p}_m| + |\Delta \vec{p}_A| + |\Delta \vec{p}_S|.$$

Here, we consider the following main channels of transfer of a momentum to particles:
- macroscopic hydrodynamic impact leading to increments of the momenta of the particles (the impact can be realized, in particular, by longitudinal acoustic waves in a medium)

$$\Delta p_m \approx m \Delta u_{sh}; \qquad (2.36)$$

- direct impact increment of a momentum in the electromagnetic field

$$(\Delta p)_A \approx qA \text{ (in IS units)}; \qquad (2.37)$$

- increment of a momentum in the field of entropy gradients

$$(\Delta p)_S \approx \hbar(\nabla S). \qquad (2.38)$$

It is clear that the action of sources of impulsive excitation on a system of many particles leads to a nonequilibrium state. As was shown in the works of A. Vlasov [13-15], the kinetic equation for the collective states of a system of particles with distribution function $f(\vec{r}, \vec{u}, t)$ can be presented in a closed divergent form in the Finsler space.

Usually even in the case of a strong deviation from the equilibrium, the system is described within the Prigogine method of locally equilibrium distributions [38] with parameters depending on the spatial coordinates. In this case, the kinetic theory describes the evolution of both a state of the system of particles and the distribution of particles in the configurational space. In this case, the sources are usually positioned on boundary of the region under consideration and act on different particles of the system differently. ***The main parameter of a driver defining the kinetics is the power flow density on the boundary of the system***.

***In the cases where a driver initiates the action of a mass force on the system,*** it renders, by definition, the practically identical action on all particles irrespective of their location in the system. In other words $\rho$ particles in unit volume receive the same momentum increment $\Delta \vec{p}$ for the time $\Delta t$. ***Therefore, the parameter defining the openness of the system is, naturally, the bulk density of a power absorbed in the system,*** $P_A = v_{eff} \rho_W$.

In the case of the action of a nonstationary vector potential, the relations $E \approx v_{eff} A$ and $\rho_W \approx (v_{eff} A)^2$ are satisfied, and we have

$$P_A \approx v_{eff} |E_A|^2 \approx v_{eff}^3 A_{eff}^2 \approx \frac{A_{eff}^2}{\tau_{eff}^3}. \qquad (2.39)$$

It is seen that the force action efficiency increases as the third power of the frequency with the corresponding decrease in the impact duration.

We now construct a dimensionless parameter characterizing the degree of "impactness" of an action. To this end, we estimate the dissipation power $P_{dis}$ and consider the dimensionless ratio $Q_{imp} = \dfrac{P_A}{P_{dis}}$ (impact factor or the coefficient of impactness) of the power density of the driver to the dissipation power density $P_{dis} \approx \dfrac{\rho T}{\tau_{dis}}$. In view of (2.32), the parameter of impactness

$$Q_{imp} = \frac{A^2}{\rho T} \frac{\tau_{dis}}{\tau_f^3} \approx \left(\frac{\tau_{dis}}{\tau_f}\right) \cdot \left(\frac{R_{WZ}}{r_e}\right) \cdot \left(\beta_T \frac{a}{a_{dis}}\right)^2, \tag{2.40}$$

where $r_e = \dfrac{e^2}{m_e c^2}$ is the classical radius of an electron, and $R_{WZ}$ is the radius of a Wigner--Seitz cell.

***The distinctive property of the parameter of impactness is its reciprocal dependence on the cube of the characteristic duration of an impact action.***

For the system of particles, the bulk density of a consumed power characterized by the parameter of impactness is the nonequilibrium source in the kinetic equation for the distribution function of particles, which is the equation of continuity in the space with coordinates $(\vec{r}, \vec{u})$ for the effective medium represented by the probability distribution:

$$\frac{\partial f(\vec{r}, \vec{u}, t)}{\partial t} + div_{\vec{r}}\left(\vec{u} f(\vec{r}, \vec{u}, t)\right) + div_{\vec{u}}\left(\langle \dot{\vec{u}} \rangle f(\vec{r}, \vec{u}, t)\right) = \Psi(r, p). \tag{2.41}$$

Here, $\langle \dot{\vec{u}} \rangle = \dfrac{\int d\dot{\vec{u}}\, \dot{\vec{u}}\, f(\vec{r}, \vec{u}, \dot{\vec{u}}, t)}{f(\vec{r}, \vec{u}, t)}$ is the acceleration averaged over the whole ensemble of particles, and the distribution function is defined in the space of support elements, the Finsler space. It will be described below. We will show that the ***properties of the space, where the distribution functions are defined, are of great importance and allow one to naturally describe the self-organization of the systems of particles even without explicit presence of forces of the fundamental nature***.

Consider a system of particles under the action of a dominating perturbation of mass forces without any dependence on the coordinates: $\Psi(r, p) \equiv \Psi(p)$.

It is clear that if all positions of particles in the system are equivalent for the action of a mass force, and if the condition of domination is satisfied, then the good zero approximation is a nonequilibrium system with spatially and statistically homogeneous properties, so that its principal evolution runs in the energetic conponent of the phase space.

***The source or sink of energy*** $\Psi(p) \approx Q_{imp} \phi(p)$ ***(mass force), which is homogeneous in the whole space, is characterized*** (by the Heisenberg uncertainty relation) by a strong localization in the momentum space. So, we may consider a linear combination of $k$ ***delta-functions as sources and sinks*** concentrated near certain points of the momentum space $p_k$.

For the isotropic part of the distribution function, it is convenient to pass from momenta to energies with the use of the dispersion law $\varepsilon = \varepsilon(p)$ and to present the kinetic equation in the form [39-40]

$$\frac{\partial f(\varepsilon, t)}{\partial t} + \frac{1}{g(\varepsilon)} \frac{\partial}{\partial \varepsilon}\left(\Pi(\varepsilon, \{f\})\right) = \sum_k \left(Q_{imp}\right)_k \frac{1}{g(\varepsilon)} \delta(\varepsilon - \varepsilon_k), \tag{2.42}$$

where we used the density of states $g(\varepsilon)$ and the flow of particles in the phase space $\Pi(\varepsilon, \{f\})$. In the Vlasov equation (2.45), the flow in the phase space corresponded to classical statistics:

$$\Pi(u, \{f\}) = \langle \dot{\vec{u}} \rangle f(\vec{r}, \vec{u}, t). \tag{2.43}$$

Let the particles of a substance be fermions. Taking the properties of quantum statistics into account, we will use, first of all, the fact that the mean acceleration of fermions $\langle \dot{\vec{u}} \rangle$

1) is proportional to the number of free sites for the evolution in the energy space,
   i.e., to the quantity $\left(1 - f(\vec{r}, \vec{u}, t)\right)$;

2) is caused by mass forces and is determined, as is seen from the analysis of the Schrödinger equation, by the entropy $S(\{f\}) \approx \ln(f)$.

In view of this, we restrict ourselves in the expansion of the acceleration $\langle \ddot{u} \rangle$ by terms up to the first derivative $\frac{\partial S}{\partial \varepsilon}$. Then the flow in the kinetic equation for fermions takes the form

$$\Pi(\varepsilon, \{f\}) = \bar{a}_0(\varepsilon)\left( T_{eff} f \frac{\partial S(\{f\})}{\partial \varepsilon} + (1-f)f \right). \tag{2.44}$$

In the regions between sources and sinks, Eq. (2.42) ensures the constancy of the flows with the corresponding sign. For the zero coefficient of impactness, the solution is the Fermi--Dirac function $f(\varepsilon) = \dfrac{1}{1 + \exp\left(-\dfrac{\varepsilon_F - \varepsilon}{T_{eff}}\right)}$. If the impactness is nonzero, the distribution function generalizes the equilibrium distribution:

$$f_q(\varepsilon) = \frac{1}{1 + \exp_q\left(-\dfrac{\varepsilon_F - \varepsilon}{T_{eff}}\right)}, \tag{2.45}$$

where the parameter of nonequilibrium $q$ is determined by the parameter of impactness, and the exponential function is replaced by the functions [36] with power asymptotics,

$$\exp_q(-x) = \left(1 + \frac{q-1}{q}(-x)\right)^{\frac{q}{q-1}}, \quad q = \sqrt{1 + \alpha_I (Q_{imp})_k}. \tag{2.46}$$

The quantity $\alpha_I$ in the formula for the parameter $q$ is determined by the information transfer rate along a communication channel between the scales and can be evaluated by the Shannon--Hartley theorem:

$$\alpha_I \approx \frac{\Delta \omega}{\omega_{eff}} \log_2(1 + Q_{imp}) \approx \frac{\xi(t)}{Q} \log_2(1 + Q_{imp}). \tag{2.47}$$

Here, $\Delta \omega$ is the transmission band of a communication channel, $\omega_{eff}$ is the effective frequency representing the action of a driver (frequency of electromagnetic signals, inverse duration of the front of pulses of the electromagnetic field, etc.), $Q$ is the quality of the oscillatory system, and the function $\xi(t)$ represents a possible modulation of information.

Using the distribution function of particles (2.40), we can write the distribution functions $f_{bq}(\varepsilon)$ over energies for holes (antiparticles):

$$f_{bq}(\varepsilon) = 1 - f_q(\varepsilon) = \frac{\exp_q\left(-\dfrac{\varepsilon_F - \varepsilon}{T_{eff}}\right)}{1 + \exp_q\left(-\dfrac{\varepsilon_F - \varepsilon}{T_{eff}}\right)}. \tag{2.48}$$

This formula can be used for the determination of the levels of fluctuations and excitation of the vacuum state of particles, for example, electrons and positrons.

Having defined the main parameters of drivers, we pass to the analysis of the general evolution of systems with constraints and to the clarification of its nature and mechanisms.

## 3. GEOMETRODYNAMICS OF SYSTEMS WITH VARYING CONSTRAINTS

The main element of kinetic theory, namely the distribution function of particles, is practically the same as the probability density distribution (the squared modulus of the wave function), which is determined by solving the Schrödinger equation. However, there exists a difference between kinetics and quantum mechanics. It consists in that ***the Schrödinger equation and other equations of quantum mechanics contain nonlocal terms*** (which is confirmed in numerous experiments), ***whereas the ordinary approaches to kinetics based on classical dynamics involve no nonlocality***.

The absence of a possibility to describe the nonlocal effects observed experimentally is the main shortcoming of practically all approaches to kinetics. A generalization of kinetics to nonlocal processes would erase these differences between classical and quantum descriptions of systems.

This purpose was attained, in the basic part, by A. Vlasov on the way of the geometrization of kinetics in the nonlocal mechanics developed by him. An analogous approach is used in our theory of evolution.

**3.1. Space-time with Finsler geometry**

The geometric interpretation of mass force action on objects of any nature is that the mass force creates a structure of space-time wherein the further evolution of the system occurs. Naturally, there is a need to consider the physical processes in Riemann spaces, the fiber spaces with different bases, Cartan and Finsler spaces, etc.

The geometries of Euclid and Riemann, which are usually applied to physics, concern only local properties. To describe the nonlocality of a system, A. Vlasov used the geometry of support elements (the geometry of a stratified space) [14-15], whose advantage consists in that the kinematic characteristics of the dynamics of particles become inherent internal characteristics of the system and are not imposed from outside. Any particle is characterized nonlocally, i.e., by the whole spectrum by the own geometric and kinematic properties for every time moment $t$: $\vec{r}, \vec{u}, \dot{\vec{u}}, \ddot{\vec{u}}, \dddot{\vec{u}}, \ldots$

The differentials of the independent coordinates, $x^0 = ct$, $x^1$, $x^2$, $x^3$, are infinitely small intervals basing on a current point $M$ in the four-dimensional Riemann space-time, i.e., in the space-time with metric properties that are determined by the metric, namely the elementary interval written in terms of the differentials of coordinates of the space:

$$ds^2 = g_{ik} dx^i dx^k, \quad i = 0,1,2,3, \quad k = 0,1,2,3. \tag{3.1}$$

Sometimes, it is convenient to separate the spatial coordinates, the time coordinate, and the spatial interval with the corresponding metric:

$$ds^2 = g_{\alpha\beta} dx^\alpha dx^\beta + 2 g_{0\alpha} dx^0 dx^\alpha + g_{00}(dx_0)^2, \quad dl^2 = \gamma_{\alpha\beta} dx^\alpha dx^\beta, \quad \gamma_{\alpha\beta} = -g_{\alpha\beta} + \frac{g_{0\alpha} g_{0\beta}}{g_{00}}.$$

The differentials lie on different surfaces and, therefore, are independent vectors. The point of the space-time and the collection of differentials of different orders (support elements) form a space with larger dimension, namely the Finsler space (the space of support elements). The kinematic quantities are expressed in terms of the corresponding differentials:

$$u^0 = \frac{dx^0}{d\tau} = c\frac{dt}{d\tau}, \quad u^\alpha = \frac{dx^\alpha}{d\tau}, \quad \dot{u}^\alpha = \frac{d^2 x^\alpha}{d\tau^2}, \quad \ddot{u}^\alpha = \frac{d^3 x^\alpha}{d\tau^3}, \ldots, \quad \alpha = 1,2,3. \tag{3.2}$$

Here, $\tau$ is the intrinsic time of particles, which is invariant relative to changes of the reference system and the laboratory time $t$. If the reference system is changed in the space-time, the values of coordinates are connected by the relations with nonzero Jacobian:

$$x^{\alpha'} = \varphi^{\alpha'}(x^0, x^1, x^2, x^3), \quad \det\left|\frac{\partial x^{\alpha'}}{\partial x^\beta}\right| \neq 0. \tag{3.3}$$

Seven degrees of freedom in the Finsler space are physically obvious: they are four coordinates of the four-dimensional space-time, $x^0, x^1, x^2, x^3$, and three velocities in the coordinate space, $u^1 = \frac{dx^1}{d\tau}$, $u^2 = \frac{dx^2}{d\tau}$, $u^3 = \frac{dx^3}{d\tau}$. The definition of 4-dimensional velocity includes a new element, the intrinsic time. Moreover, the eighth degree of freedom with dimension of velocity, $u^0 = c\frac{dt}{d\tau}$, appears.

Since the eighth degree of freedom $u^0$ is directly related to the flow of the physical time inside the ensemble of interacting particles, it is natural to assume the connection of this eighth degree of freedom with the physical properties of irreversibility of the system, degree of its "openness," and flows of the entropy in the system of particles. This point will be clarified in the subsequent study of the connection between the geometry and the processes of evolution.

The curvature of the stratified space-time is defined by the coherent acceleration of the system (and, hence, by mass forces). The category of motion of particles is included in the space of support elements on the same primary level as the category of space-time. Moreover, the forces are considered as a factor forming the properties of the space-time and the possible motions, which are already connected continuously with the image of a particle.

The ordinary phase space differs from the space of support elements in the following. The velocities in the phase space occupy the whole region in a vicinity of the corresponding point of the coordinate space, whereas the velocities in the space of support elements are in the plane tangent to the world lines passing through the given point of the space-time and are obtained by the differentiation along the world lines of particles.

This results in that the velocities in the Finsler space of support elements, as distinct from the Riemann space, are transformed always by a linear law, even for the arbitrary nonlinear transformations of coordinates (3.3).

In other words, the 8-dimensional Finsler space with coordinates $\left(x^0, x^1, x^2, x^3, u^0, u^1, u^2, u^3\right)$ is characterized by the transformations

$$x^{\alpha'} = \varphi^{\alpha'}\left(x^0, x^1, x^2, x^3\right), \quad u^{\alpha'} = u^\alpha \frac{\partial x^{\alpha'}}{\partial x^\alpha}. \tag{3.4}$$

Elements $a^\alpha = a^\alpha\left(x^\sigma, u^\sigma\right)$ form a contravariant vector, if they are transformed as the vector of velocities $u^\alpha$ (see (3.4))

$$a^{\alpha'} = a^\alpha \frac{\partial x^{\alpha'}}{\partial x^\alpha}, \tag{3.5}$$

and $a_\beta = a_\beta\left(x^\sigma, u^\sigma\right)$ form a covariant vector, if they are transformed with the help of the relations

$$a_{\beta'} = a_\beta \frac{\partial x^\beta}{\partial x^{\beta'}}. \tag{3.6}$$

*According to the Hausdorff theorem of the metrics of topological spaces, the general metric of the space $\tilde{ds}^2 = ds^2 + ds_u^2$ can be set by a sum of the independent metrics of the space-time $ds^2$ and the metric of the tangent bundle, i.e., that of the space of velocities $ds_u^2 = q_{ik} du^i du^k$.*

In this case, the metric of the space-time part can be of two basically different types:

1) metric tensor of the space-time depends only on coordinates and the time: $g_{ik} = g_{ik}\left(x^l\right)$;

2) metric tensor of the space-time depends on velocities (and, possibly, accelerations) implicitly, as on parameters: $g_{ik} = g_{ik}\left(x^l, \{\vec{u}, \dot{\vec{u}}, \ddot{\vec{u}}, \dddot{\vec{u}}, ...\}\right)$.

In the first (isotropic) case corresponding to a weak deviation of the collective system from the equilibrium without rearrangement of the internal structure, the metric depends only on coordinates.

In the second case, we have an anisotropic scenario corresponding to the collective system with self-organizing internal structure. The dependence of the metric coefficients on velocities (and/or accelerations) leads to a specific anisotropy of the space-time: *in any infinitely small region, the space-time is anisotropic, and its properties depend on the direction of motion and the acceleration of particles.*

*Namely the anisotropy of the space-time, which arises obviously at the coherent acceleration of the system, is the main reason for the formation of macroscopic quantum (coherent, nonlocal) objects of the shell type.*

The particles of a shell, which are organized in a collective state, i.e., MQO, form a noninertial reference system. The acceleration of this collective reference system is reflected in the space-time curvature resulting in the difference between the intrinsic and laboratory times, which can cause, as will be shown below, the explosive change of space-time scales.

### 3.2. Geodesic lines in an evolving system

Under the action of mass forces on a system of particles, the particles are identically accelerated and form a noninertial system. In a vicinity of the arbitrary point of the space-time, the dynamics of the system is set by covariant accelerations and, hence, by covariant derivatives of the velocity. In turn, the covariant derivatives of the velocity are given by the tensor of accelerations. As follows from the Gauss principle and the principle of dynamical harmonization, the constraints in the system are changed with the help of a variation of accelerations. Therefore, the constraints between kinematic quantities and the limitations imposed on them are determined by coherent accelerations. It is clear that, in this case, one of the significant quantities is the covariant velocity differential $D_\beta u^\alpha$, whose value is set by the tensor of accelerations $a^\alpha_\beta(x^\sigma, u^\sigma)$ and is determined by the structure of constraints:

$$D_\beta u^\alpha = \frac{\partial u^\alpha}{\partial x^\beta} + C^\alpha_{\beta\gamma} u^\gamma = a^\alpha_\beta(x^\sigma, u^\sigma). \tag{3.7}$$

Here, $C^\alpha_{\beta\gamma}$ are the generalized coefficients of connectedness, which coincide with the Chistoffel symbols $\Gamma^\alpha_{\beta\gamma} = \frac{1}{2} g^{\alpha\sigma} \left( \frac{\partial g_{\beta\sigma}}{\partial x^\gamma} + \frac{\partial g_{\sigma\gamma}}{\partial x^\beta} - \frac{\partial g_{\beta\gamma}}{\partial x^\sigma} \right)$ in the simplest case where $a^\alpha_\beta(x^\sigma, u^\sigma) = 0$ and can be expressed in terms of the space-time metric. Let us now calculate the covariant acceleration by using the covariant velocity differential:

$$\frac{Du^\alpha}{d\tau} = \frac{(D_\beta u^\alpha) dx^\beta}{d\tau} = a^\alpha_\beta \frac{dx^\beta}{d\tau}. \tag{3.8}$$

If $a^\alpha_\beta \frac{dx^\beta}{d\tau} \neq 0$, then there exists a nonzero external covariant acceleration, and the system undergoes the action of external forces. But the situation where $a^\alpha_\beta \frac{dx^\beta}{d\tau} = 0$ can be also realized; i.e., the covariant acceleration of the system is zero, and no external forces affect the system. However, the coherent acceleration can be present, nevertheless, inside the system, the constraints, $a^\alpha_\beta(x^\sigma, u^\sigma) \neq 0$, can hold, and the influence of the motion of particles on the space-time metric can be revealed in a change of both spatial and temporal scales.

Let us set the tensor of accelerations in the simplest form:

$$a^\alpha_\beta = a \left( \delta^\alpha_\beta - \frac{u^\alpha u_\beta}{c^2} \right), \quad u^\alpha u_\alpha = c^2, \quad a = const. \tag{3.9}$$

In this case, the force action on the system is absent. Indeed, for tensor (3.9), we have

$$a^\alpha_\beta \frac{dx^\beta}{d\tau} = a \left( \delta^\alpha_\beta - \frac{u^\alpha u_\beta}{c^2} \right) \frac{dx^\beta}{d\tau} = a \left( \delta^\alpha_\beta - \frac{u^\alpha u_\beta}{c^2} \right) u^\beta = a \left( u^\alpha - \frac{u^\alpha u_\beta u^\beta}{c^2} \right) = 0. \tag{3.10}$$

Relations (3.7) are 16 equations for four unknowns $u^\alpha$; i.e., these equations have no solutions in the general case with arbitrary generalized coefficients of connectedness $C^\alpha_{\beta\gamma}$. The condition of existence of solutions of Eqs. (3.7) imposes certain limitations on the quantities $C^\alpha_{\beta\gamma}$ or $\Gamma^\alpha_{\beta\gamma}$ and, hence, on the metric.

The dynamics of particles occurs along the geodesic lines in the Finsler space, whose geometry varies in the general case in correspondence with the running evolution of the internal structure of particles and the system on the whole according to the equations

$$\frac{d^2 x^i}{dt^2} + C^i_{jk} \frac{dx^j}{dt} \frac{dx^k}{dt} + F^i = 0. \tag{3.11}$$

E. Cartan and J. Schouten showed that, by the differential equations (3.11), it is possible to restore the geometry, i.e., the metric. Conditions (3.8) are consistent with a nonstationary metric:

$$ds^2 = (dx^0)^2 - \sigma^2(x^0) g_{\alpha\beta}(x^1, x^2, x^3) dx^\alpha dx^\beta, \quad \sigma(x^0) = exp_q\left(\frac{x^0}{c\tau_{eff}}\right), \quad \alpha, \beta = 1, 2, 3, \tag{3.12}$$

where $g_{\alpha\beta}(x^1, x^2, x^3)$ is the spatial part of the Riemann metric. In this metric,

$$\kappa_{10,10} = \frac{a^2_{cog}}{c^4} exp_q\left(\frac{x^0}{c\tau_{eff}}\right) \quad \text{and} \quad \kappa = 2\frac{a^2_{cog}}{c^4} \tag{3.13}$$

are the main independent component of the curvature tensor and the scalar curvature of the space-time, which depend on the parameter $q$ closely connected with the order parameter and the coefficient of impactness of a driver.

As was shown in Section 2, the controlled change of the entropy of the system can be an efficient driver. In other words, it can induce a coherent acceleration of the system and, hence, the space-time curvature. The strongest changes of the entropy initiate the explosive processes of growth or decay of structures. Let us consider an important example of the explosive clusterization in the system of monomers (see [1,5]) and calculate the entropic forces and the space-time curvature created by the process of clusterization.

The system of particles aggregating as a result of binary contacts is a set of clusters of various sizes. The distribution of clusters over sizes, i.e., the concentration of clusters with size $k$ (clusters consisting of $k$ monomers) as a function of the time is described by the system of reactions

$$A_{k0} + A_{k0} \to A_{2k0}, \quad A_{k0} + A_{2k0} \to A_{3k0} \ldots$$

In this case, the equation for the concentrations $C_k$ of clusters including $k$ monomers can be written in the form of the Smoluchowski coagulation equation [41]. This equation involves the competition of two processes: the sticking of a cluster with monomers, i.e., the increase of the size of a cluster, and the breaking of clusters, i.e., the growth of the number of clusters with low masses. For the probability $K_{ij}$ of the sticking of clusters with sizes $i$ and $j$, we take the approximation such that this probability is proportional to the product of the surface areas of the input clusters: $K_{ij} \propto (ij)^{2/3}$. According to the solution of the Smoluchowski equation, the time dependence of the mean size $s(t)$ of a cluster manifests the explosive behavior:

$$s(t) = \frac{s_0}{\left(1 - \dfrac{t}{t_c}\right)^{1/6}}. \tag{3.14}$$

Here, $t_c$ is the time moment of the phase transition into the state of a global cluster. By the order of magnitude, this time is equal to several collision times. The explosive growth (3.14) of a cluster is directly related to the change of the number of constraints in the system and, hence, to the change of the entropy. The particles, which are organized into the collective state (i.e., MQO), form the own reference system. The acceleration of this collective reference system arosen due to the action of a mass force is reflected in the space-time curvature and, in particular, in the difference between the intrinsic and laboratory times. In this case, the local time and its intervals differ from the corresponding values in the laboratory reference system in agreement with metric (3.12):

$$\tau/t_0 = ln\left(1 + (1-q)\left(\frac{x^0}{c\,t_0}\right)\right)^{\frac{1}{1-q}} = ln\left(exp_q\left(\frac{x^0}{c\,t_0}\right)\right). \tag{3.15}$$

The parameter $q$ is connected with the order parameter $\eta$ with the help of relations obtained in [1]. It follows from formula (3.15) that the laboratory and intrinsic times coincide for the nonequilibrium parameter equal to 1. As the degree of nonequilibrium and the deviation of the parameter $q$ from 1 increase, the intrinsic time rapidly decelerates or accelerates as compared with the laboratory one, by depending on the sign of the acceleration of the nonequilibrium reference system (NRS) (and, respectively, on the direction of the deviation of $q$ from 1).

Dependence (3.15) of the ratio of the intrinsic time to the laboratory one on the laboratory time for various values of the parameter of nonequilibrium $q$ is presented in Fig. 3.1.

The variation of the intrinsic time as compared with the laboratory one leads, in turn, to a decrease of the light velocity in the region occupied by the growing cluster. The decrease of the light velocity $c = \dfrac{c_0}{\sqrt{n}}$ manifests itself as the effect of increasing the refractive index $n = \dfrac{1}{\left(ln\left(exp_q\left(\dfrac{x^0}{c\,t_0}\right)\right)\right)^2}$ near the growing cluster. It is worth noting that the described effect of increasing the refractive index and decreasing the light velocity in a vicinity of growing crystals was discovered, in fact, experimentally as earlier as 1902 and was described in work [25].

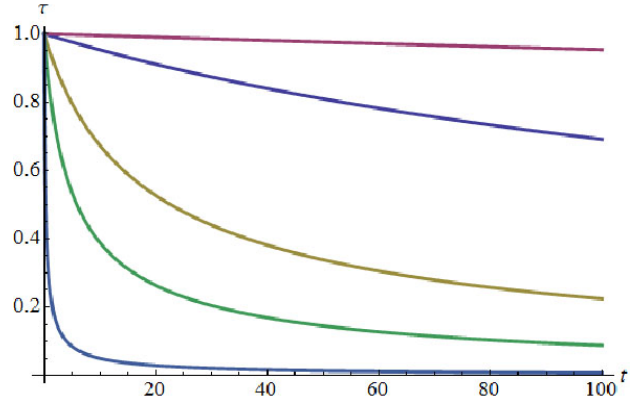

*Fig. 3.1. Ratio of the intrinsic time to the laboratory one versus the laboratory time for values of the parameter $q = 1.001, 1.01, 1.1, 1.30, 1.90$. Curves correspond to values increasing from top to bottom. For $q = 1$, the laboratory and intrinsic times coincide, and their ratio is equal to 1.*

We note that the geodesic world lines of particles, along which the particles move in correspondence with the principle of dynamical harmonization, are, in fact, the characteristics of a kinetic equation of the Vlasov type. This allows us to write the very kinetic equations and to realize the connection between the dynamical and statistical descriptions of the evolution of systems with varying constraints. Below, we will analyze the solutions of covariant kinetic equations, which allow one to answer the majority of questions posed by the "shell" model of self-organization (see [1]) and to obtain the equations of dynamical harmonization for it.

**3.3. Covariant kinetic equations for particles and their solutions**

In order to describe the many-particle interactions, the mean accelerations of particles $\langle \ddot{\vec{u}} \rangle = \frac{1}{m} \vec{F}_B$ in the Vlasov equation (2.36) are usually determined by self-consistent electromagnetic fields. As follows from Section 2, the analysis of the evolution of MQO should consider not only the fields of fundamental interactions, but also the entropic forces, which modify, in fact, the Vlasov equation, supplementing it by the collision integral in the divergent form:

$$\frac{\partial f(\vec{r},\vec{u},t)}{\partial t} + div_{\vec{r}}\left(\vec{u}\, f(\vec{r},\vec{u},t)\right) + div_{\vec{u}}\left(\langle \ddot{\vec{u}} \rangle f(\vec{r},\vec{u},t)\right) = div_{\vec{u}}\left(\vec{j}_S\right);$$

$$\vec{j}_S = \left(-\frac{1}{m}\nabla_r\left(\omega_{e\!f\!f} S_q\right)\right) f(\vec{r},\vec{u},t);\ \langle \ddot{\vec{u}} \rangle = \frac{1}{m}\vec{F}_B,\ \omega_{e\!f\!f} \approx \frac{2\pi}{\tau_{e\!f\!f}}. \tag{3.16}$$

The action of a mass entropic force on the system of particles and the continuously related acceleration compel the system to reconstruct its internal structure and, thus, to evolve in a tangent bundle of the space-time according to the variational principle of dynamical harmonization.

A change of the structure causes variations of the distribution functions of particles of the system, which are related to the processes of localization or delocalization and, in turn, affect the dynamics of particles through the entropic forces. The interrelation of the evolution of a system in the corresponding layer and the four-dimensional basis of the Finsler space-time is realized through the accelerations and the distribution functions.

The kinetic equation (3.16) for the particles composing MQO can be represented in the 8-dimensional space of support elements in the covariant form

$$\widehat{Div}_r\left(\vec{u}\, f\right) + div_u\left(\left\langle \frac{D\vec{u}}{d\tau} \right\rangle f\right) = 0. \tag{3.17}$$

Since

$$\widehat{Div}_r\left(\vec{u}\, f\right) = u^\alpha \widehat{D}_\alpha f + f \widehat{D}_\alpha u^\alpha,\ div_u\left(\left\langle \frac{D\vec{u}}{d\tau} \right\rangle f\right) = \frac{\partial}{\partial u^\alpha}\left(\left\langle \frac{D\vec{u}}{d\tau} \right\rangle^\alpha f\right),$$

$$\widehat{D}_\alpha f = \frac{\partial f}{\partial x^\alpha} - \Gamma^\sigma_{\alpha\gamma} u^\gamma \frac{\partial f}{\partial u^\sigma},\ \Gamma^\sigma_{\alpha\gamma} = \frac{1}{2}g^{\mu\sigma}\left(\frac{\partial g_{\mu\alpha}}{\partial x^\gamma} + \frac{\partial g_{\mu\gamma}}{\partial x^\alpha} - \frac{\partial g_{\alpha\gamma}}{\partial x^\mu}\right),$$

We can write (3.53) by components:

$$u^\alpha \frac{\partial f}{\partial x^\alpha} - \Gamma^\sigma_{\alpha\gamma} u^\gamma \frac{\partial f}{\partial u^\sigma} + \frac{\partial}{\partial u^\alpha}\left(\left\langle\frac{\widehat{D}\vec{u}}{d\tau}\right\rangle^\alpha f + P_s\right) = 0. \qquad (3.18)$$

Let us analyze the solutions of the kinetic equation in a significant partial case where the explicit contribution of the divergence $div_{\vec{u}}(.)$ to the kinetic quations can be neglected. Then the covariant equation of quasistationary states takes the form

$$u^\alpha \frac{\partial f}{\partial x^\alpha} - \Gamma^\sigma_{\alpha\gamma} u^\gamma \frac{\partial f}{\partial u^\sigma} = 0. \qquad (3.19)$$

This approximation is valid in two cases:
- if the external forces are completely absent, and the flow in the phase space, $P_S = 0$ (this case corresponds to the full equilibrium of the system);
- if the external forces do not act directly in the system, but the flows of energy, particles, or the entropy are constant in the phase space, $P_S = const$ (this case corresponds to a strong deviation from the equilibrium).

In the nonequilibrium case, the kinetic equations for systems with varying constraints have quasipower and power solutions [39-40]. In this case, the exponent of a solution depends on the flows created in the system [40], the external forces inside the dynamical system of particles can be neglected, and the whole action of mass forces is determined by the entropy flows that are generated on the boundary of an MQO-shell and disappear in the orthogonal direction, where the delocalization and the growth of a structure in the system occur.

Consider the solutions of the kinetic equation (3.19), which are isotropic in the space of velocities and stationary in the laboratory time, i.e., we assume that

$$\partial f / \partial x^0 = 0;\ \partial g_{\alpha\beta}/\partial x^0 = 0,\ g_{0i} = 0,\ i = 1,2,3.$$

We emphasize that this stationary state does not assume the independence of the intrinsic time. In the indicated approximation, we have

$$\Gamma^0_{00} = \Gamma^0_{ik} = 0;\ \Gamma^k_{i0} = 0;\ \Gamma^0_{0i} = \frac{1}{2g_{00}}\frac{\partial g_{00}}{\partial x^i};\ \Gamma^i_{00} = -\frac{1}{2}g_{ik}\frac{\partial g_{00}}{\partial x^k}.$$

Separating the variables, we present the distribution function in the form

$$f\left(x^\alpha, u^\alpha, u^0, t\right) = f\left(x^\alpha, u^\alpha, u^0\right) = \rho\left(x^\alpha\right)\psi\left(u^2\right)\psi_0\left(u^0\right), \qquad (3.20)$$

$$(1-q)\xi^2 = g_{\alpha\beta}\xi^\alpha\xi^\beta,\ \xi^\alpha = \frac{u^\alpha}{\sqrt{-(1-q)g_{oo}}u^0},\ u^2 = u_\alpha u^\alpha. \qquad (3.21)$$

Substituting (3.21) in (3.20) and (3.19) and separating the variables, we obtain the system of equations, which is exactly solvable:

$$\rho\left(x^i\right) = \rho_0 exp_q\left(-\frac{U(x)}{w(q)}\right),\ \psi_0\left(u^0\right) = \psi_0(0)\frac{1}{u_0^{\frac{q_{cr}}{q_{cr}-1}}},\ \psi\left(\xi^2\right) = \left(exp_q\left(-\xi^2\right)\right)^q. \qquad (3.22)$$

The obtained solutions reflect the fact that, in the absence of the entropy flow (i.e., for $q = 1$), the homogeneous equilibrium case is realized, since the distribution function $\psi(\xi^2)$ over velocities (or energies) transits in the Maxwell distribution function, the distribution function $\rho(x^i)$ becomes the Boltzmann distribution, and the function $\psi_0(u^0)$ is constant, so that there is no difference between the local and laboratory times. If the entropy flows are present in the system ($q \neq 1$), the exponential distribution functions over energies and coordinates become the quasipower ones.

### 3.4. Anisotropy of states in a noninertial dynamical system

The action of the entropic fields of mass forces initiating the coherent acceleration of particles transforms the whole system into a noninertial reference system (NRS). The inertial reference systems (IRS) are associated with the absence of flows and coherent accelerations in the reference system, i.e., with equilibrium systems without evolution. *On the contrary, namely the action of mass forces on IRS and its transformation in NRS with some coherent acceleration and flows in it compel the particles to the evolution, i.e., to a change of the internal structure of constraints between particles, energy of constraints, and mass defect of the system.*

In the general case, the system (like a spheroid), which is isotropic at the initial time moment and has a spatial distribution of particles with characteristic scale $l_0$ in the inertial system, evolves in the noninertial system in a deformed anisotropic state with the number of external space scales more than 1 and becomes similar to an ellipsoid-"pancake". In a sufficiently general case, we may distinguish two basically different orthogonal directions:
- along the direction of the acceleration;
- in the plane orthogonal to the acceleration.

By controlling the anisotropy of the wave function of a quantum system, it is possible to control the localization of the system and, hence, the probabilities of the processes of evolution related to the overcoming of energy barriers. This is obvious even at the analysis of pure states.

The Heisenberg uncertainty relation for one degree of freedom takes the form $\Delta x \Delta p_x \geq \frac{\hbar}{2}$. In the complete phase space (e.g., for three degrees of freedom), this relation determines the size of its minimal cell:

$$\Delta \Omega_{ph} \geq \frac{\hbar^3}{8}. \tag{3.23}$$

In the simplest case, we will take the anisotropy of the evolution into account, by introducing two scales as macroscopic geometric characteristics of the system instead of its single radius:
- $l_- < l_0$ in the direction of the "flattening" of the system;
- $l_+ > l_0$ in the orthogonal directions, along which the "flattening" happens.

The less scale $l_-$ can be called a scale of spatial coherence of the system, which characterizes the "pancake" thickness. The larger scale $l_+$ is a characteristic scale of the interaction, which characterizes the maximal length of correlations in the system.

Consider the case of a strong deviation from the local equilibrium. Though the external forces do not act directly in the system, the entropy flow is constant in the phase space, and the density distribution in the bounded system and the erosion of the boundary happen in agreement with a distribution function of the type (3.22), rather than with equilibrium distribution functions.

The density distribution in the system with mass number $A$ that consists of monomers with mass number and characteristic size $l_{str}$ is described at a distance $\Delta r$ from the center by the squares of the corresponding wave functions with characteristic scale $l_0(\eta) = l_{str} \left( \frac{A}{A_{str}} \right)^{\frac{\eta+1}{3+\eta}}$:

- $\exp\left(-\left(\frac{\Delta r}{l_0}\right)^2\right)$ for the equilibrium noncoherent part of the system;
- $\exp_q\left(-\left(\frac{\Delta r}{l_0(q)}\right)^2\right)$ for the coherent part of the system with regard for the action of entropic forces.

Without the entropy flow (for $q=1$), a homogeneous equilibrium state is realized. In this case, the distribution function over velocities (or energies) transits in the Maxwell distribution. If the entropy flows are present in the system ($q \neq 1$), the exponential distribution functions over energies and coordinates becomes quasipower one.

The processes of self-organization essentially depend on the direction of the entropy flow, since there exist two types of behavior of the distribution function: localization and delocalization. They correspond to different characters of the behavior of the function $\exp_q\left(-\left(\frac{\Delta r}{l_0(q)}\right)^2\right)$ for $q \leq 1$ and $q > 1$. The parameter $q$ is determined from (2.41)-(2.42) and is connected with the order parameter $0 \leq \eta \leq 1$ by the relations

$$q(\eta) = \begin{cases} q_- = 1 - \eta, & q \leq 1 \\ q_+ = \dfrac{1}{1-\eta}, & q > 1 \end{cases}. \tag{3.24}$$

As the order parameter increases, the character of decay of the wave functions in the space passes from exponential to power (see 3.22). The dependences of the scales on the order parameter in the directions of delocalization and localization take the form

$$l_+ \approx l_0 \frac{1}{\left(1 - \dfrac{\eta}{\eta_{\max}}\right)^{\gamma_{str}}}, \quad l_- \approx l_0\left(1 - \frac{\eta}{\eta_{\max}}\right), \quad \gamma_{str} \approx 1.83, \quad \eta_{\max} = 0.5. \tag{3.25}$$

The phase volume is a product of volumes in the coordinate and momentum spaces, and the volume in the coordinate space is a product of volumes in the coherent direction $\Delta\Omega_-$ and in the direction $\Delta\Omega_+$ orthogonal to it. Thus, we have $\Delta\Omega_{ph} = (\Delta\Omega_+ \Delta\Omega_-)\Delta\Omega_p$.

The volume in the plane orthogonal to the direction $l_-$ can be estimated as $\Delta\Omega_+ \approx \pi l_+^2$. We consider that the minimal size of a cell $l_-$ is attained in the coherent state: $(l_-)_{\min} \approx \dfrac{\hbar}{2p_f}$. Since the volume in the momentum space $\Delta\Omega_p \approx \dfrac{4}{3}\pi p_f^3$, we have

$$l_+ \geq \frac{1}{2\pi}\left(\frac{\hbar}{p_f}\right)^{3/2}\sqrt{\frac{3}{8l_-}}. \tag{3.26}$$

These relations are valid, if there are no correlations between coordinates and momenta. However, if the system of particles turns out in the state with coherent acceleration, then, as is seen from the above discussion, the space-time metric is changed. Hence, the geodesic lines and the dynamics of particles are changed as well. It is clear that, in this case, due to the anisotropic space-time curvature (and to the coherent acceleration), the strong correlations appear:

$$r_{xp} \approx \eta^{2(\gamma_{str}-1)/3}. \tag{3.27}$$

In the general form, the correlations $r_{xp}$ between momenta and coordinates were considered by Robertson and Schrödinger [42], who wrote the uncertainty relation in the form

$$\Delta x \Delta p_x \geq \frac{\hbar}{2\sqrt{1-r_{xp}^2}}. \tag{3.28}$$

Let us return to relation (3.23). With regard for the correlations $r_{xp}$, we can transform it into the form similar to (3.26):

$$l_+ \geq l_- \frac{1}{2\pi}\sqrt{\frac{3}{8}}\left(\frac{\hbar}{p_f l_-}\frac{1}{\sqrt{1-r_{xp}^2(a_{cog})}}\right)^{3/2} \tag{3.29}$$

As was shown in [1], this relation at the coherent acceleration of the system of particles yields the explosive delocalization of a state of the system in the direction orthogonal to the direction of the acceleration. As the shell thickness decreases, the energy becomes quantized in the direction perpendicular to the shell surface. In other words, the dispersion of momenta of particles around the momenta localizing themselves is decreased. In this case, the dispersion of coordinates of particles along the surface is sharply increased.

It was shown in [43] that, on the basis of estimates (3.27), it is possible to develop a model of the overcoming of barriers by an oscillator located in a nonstationary external field ensuring the growth of correlations.

## 3.5. Equations of dynamical harmonization of a system with varying constraints and the geometry of a stratified space-time

The most general representation of the laws of dynamics and evolution of the systems of particles is given by the variational principle of dynamical harmonization [1], which is a generalization of the Gauss and Hertz principles for the systems with varying internal structure and binding energy. It assumes that the self-organization of a system occurs as a result of the variation of the structure of constraints between its particles (elements) as a response to their coherent acceleration.

By the Gauss principle, those positions that will be occupied by the points of the system at the time moment $t + \tau$ in their real motion are distinguished between all positions admissible by constraints by the minimal value of compulsion measure $Z_G = \sum_{i=1}^{N} m_i s_i^2$ (here, $s_i(\delta a)$ is the length of a vector between the points representing the true and any possible positions of a point; it depends only the acceleration variation $\delta a$).

The optimal variations of accelerations, as was shown by Hertz, correspond to the minimal curvature of the trajectories of particles. This means that the dynamics of particles is realized along the geodesic lines corresponding to definite constraints. The notion of motion includes also a rearrangement of both the structure of the system and the field of constraints of its structural elements. While the system moves, its fractal dimension $D_f$ and binding energy $B(D_f)$ [1], which are determined by the packing of monomers composing the system, are changed.

Changes of the structure and constraints in the system vary obviously the masses of the system and its components (i.e., the inertia or sensitivity of the system relative to the external gorces acting on it).

As was shown in [1], the evolution of an internal structure of the system is determined by the principle of dynamical harmonization, which involves the possibility of a change of constraints in the system:
*under the action of external, $F_i$, and mass, $F_{str}$, forces, the system varies its trajectory and the structure in order to be consistent with the external medium and external actions, by minimizing the generalized compulsion function,*

$$Z_{dh} = \sum_{i=1}^{N} \left( m_i(D_f) w_i - \left( F_i + (F_{str})_i \right) \right)^2, \quad m_i(D_f) = \left( m_{i0} - \delta m_i(D_f) \right), \quad \delta m_i(D_f) = \frac{B_i(D_f)}{c^2}, \quad (3.30)$$

*with regard for the variations of all constraints in the system (respectively, with regard for the variation of the binding energy $B_i(D_f)$).*

*In other words, the system tends to make the trajectories of its compulsory motion under the action of mass forces to be maximally close to the trajectories of the own nonperturbed motion.*

Since a change of the internal structure of the system is regularly related to a change of its mass $m_i(D_f)$, the processes accompanied by a change of the structure are most efficient at the evolution of the system, because they can serve as both a source of energy and a means of its accumulation for the very evolution.

*It is obvious that the control over a system on the basis of the laws of evolution of its constraints (the principle of dynamical harmonization of the systems with varying constraints) is the unique efficient way o the realization of desired transformations in the system due to the use of its internal energy resources, rather than due to the direct "violence" with the use of only the external energy.*

The tool to initiate the processes of self-organization of a structure of constraints in the system is a general dominating perturbation specially selected for the given system and the appropriate coherent acceleration of the ensemble of particles composing the system.

Because a change of the structure is continuously connected with changes of the entropy and the information, the principle of dynamical harmonization describes simultaneously the targeted exchange by information and the entropy between the system and the environment. This means that the space-time geometry (curvature) and the evolution of an internal structure of the system of particles are indivisible. Such a situation is a natural continuation of properties arising in the dynamical systems with varied constraints between elements of the system under the optimization of their control.

In this case,
- the state of the system is set by a vector in the configurational space of a dynamical system,
- the constraints are set by a matrix of the constraint coefficients,

- the control is realized by the external vector of control, being the vector of forces acting on the appropriate components of the system.

An analogous situation arises also at the evolution of the system represented by a collection of monomers:
- the state of such a system is set by the positions of particles in the four-dimensional space-time and their velocities, which are tangent to the trajectories of particles at the given point and, hence, belong to a tangent bundle of the space-time;
- the constraints between monomers are characterized by their energies depending on the structure of the system, which is characterized, in turn, by the dimension (the dimension of a structure of constraints in the system) and the entropy (information);
- the evolution of the system occurs in the tangent bundle of the space-time and is governed by the equations of dynamical harmonization in a noninertial reference system with given coherent acceleration. The evolution forms the entropic forces that define the dynamics of the system of particles with varying constraints in the space-time. The coordinates in layers are the accelerations of all orders; additionally, we have a layer with the fractal dimension of a system of constraints (or with their entropy) as a coordinate;
- the dynamics of the system of particles occurs in the anisotropic space-time with a curvature that is determined by the acceleration of the noninertial reference system depending on the entropic forces;
- the control is realized by the external vector of control, which sets the contributions to the appropriate components of coherent accelerations of the system.

As was shown in the previous section, в предыдущем разделе, the action of mass forces on the system causes the rapid (as compared with a quasiequilibrium case) "flattening" of the distribution function, which corresponds to the presence of the negative flows of entropy in the system (or, what is the same, the flow of information in the system), ensures an increase of the volume, and, by this, modifies the dynamics of scales. In the general case, as the order parameter increases and the fractal dimension varies, let the localization scale $l_-$ be decreased, and let the scale $l_+$ be increased as compared with the equilibrium values by the relations

$$l_- = g_-(D_f, \delta); \quad l_+ = g_+(D_f, \delta). \tag{3.31}$$

The estimation of these functions was mase above (see (3.25)). In agreement with the principle of dynamical harmonization, the equations of evolution are determined by a minimum of the dynamical harmonization functional $Z_{dh}$ at the variation of the accelerations of the scales of localization and delocalization of the system (respectively, $w_-$ and $w_+$):

$$Z_{dh} = \frac{1}{2}(m\,w_+ - F_+)^2 + \frac{1}{2}(m\,w_- - F_-)^2 \, , \, m = m_0 - B_A(\eta, \delta)/c^2. \tag{3.32}$$

The variation by Gauss assumes the tangent plane to a current point on the trajectory of a particle to be fixed, and the transition from the dynamics in the coordinate space to that in the Finsler space is very simple. Then the variations by Gauss look as those in a tangent plane with second-order tangency at the fixed plane with first-order tangency. *The variations of accelerations (i.e., of vectors in the corresponding different planes) of all orders are independent. Therefore, the variations by Gauss lead to that the relations for the variations of accelerations are similar to those for the variations of the corresponding coordinates. Hence, the below-presented relations for accelerations do not include the first derivatives of the constraint equations*:

$$w_1 = \frac{d^2}{dt^2}l_1 = \gamma_{11}\ddot{D}_f + \gamma_{12}\ddot{\delta} \, , \, \gamma_{11} = \frac{\partial^2 g_1}{\partial^2 D_f} \, , \, \gamma_{12} = \frac{\partial^2 g_1}{\partial^2 \delta}. \tag{3.33}$$

$$w_2 = \frac{d^2}{dt^2}l_2 = \gamma_{21}\ddot{D}_f + \gamma_{22}\ddot{\delta}, \, \gamma_{21} = \frac{\partial^2 g_2}{\partial D_f^2} \, , \, \gamma_{22} = \frac{\partial^2 g_2}{\partial^2 \delta}.$$

Here,
 index 1 corresponds to the direction of delocalization $x_+$ and the scale of delocalization $l_+$,
 index 2 corresponds to the direction of localization $x_-$ and the scale of localization $l_-$.

Substituting the formulas for the accelerations in $Z_{dh}$, we obtain the dynamical harmonization functional depending on the accelerations of the fractal dimension and the deformations of scales:

$$Z_{dh}(\ddot{D}_f, \ddot{\delta}) = \frac{1}{2}\left(\gamma_{11}\ddot{D}_f + \gamma_{12}\ddot{\delta} - \frac{F_1(D_f, \delta)}{m(D_f, \delta)}\right)^2 + \frac{1}{2}\left(\gamma_{21}\ddot{D}_f + \gamma_{22}\ddot{\delta} - \frac{F_2(D_f, \delta)}{m(D_f, \delta)}\right)^2. \quad (3.34)$$

The condition of minimum of the dynamical harmonization functional with respect to the accelerations of the fractal dimension and the deformations of scales ($\frac{\partial Z_{dg}(\ddot{D}_f, \ddot{\delta})}{\partial \ddot{D}_f} = 0$, $\frac{\partial Z_{dg}(\ddot{D}_f, \ddot{\delta})}{\partial \ddot{\delta}} = 0$) leads to the system of differential equations determining the evolution of a dymanical system with varying constraints:

$$\ddot{D}_f = \frac{a_{22}G_1 - a_{12}G_2}{a_{11}a_{22} - a_{12}a_{21}}; \quad \ddot{\delta} = \frac{-a_{21}G_1 + a_{11}G_2}{a_{11}a_{22} - a_{12}a_{21}}, \quad (3.35)$$

$$a_{11} = ((\gamma_{11})^2 + (\gamma_{21})^2), \ a_{12} = (\gamma_{11}\gamma_{12} + \gamma_{21}\gamma_{22}), \ G_1 = \gamma_{11}\frac{F_1}{m} + \gamma_{21}\frac{F_2}{m}, \quad (3.36)$$

$$a_{21} = (\gamma_{11}\gamma_{12} + \gamma_{21}\gamma_{22}), \ a_{22} = ((\gamma_{12})^2 + (\gamma_{22})^2), \ G_2 = \gamma_{12}\frac{F_1}{m} + \gamma_{22}\frac{F_2}{m}.$$

The obtained equations for the order parameter and a deformation of the probability density distribution of the system and the very variational principle of dynamical harmonization, from which the equations are deduced, are the basis основой of the theory of self-organization of systems with varying constraints.

In the simple situation where the deformation varies much more slowly that the internal structure, we may consider the deformation to be given. The equation describing the evolution of a structure, which has always time to tune itself to a given deformation, was obtained in [1]. In this case, the order parameter and the fractal dimension evolve according to the equation of dynamical harmonization, which has form of the Lagrange equation describing a change of the structure of the system with the use of the corresponding Lagrange function $L_{str}$:

$$\frac{d}{dt}\left(\frac{\partial L_{str}}{\partial \dot{D}_f}\right) - \frac{\partial L_{str}}{\partial D_f} = 0, \ L_{str} = m_{str}(D_f)R_0\frac{\dot{D}_f^2}{2} + sB_A(Z, D_f)A - U_{str}(D_f). \quad (3.37)$$

Here, $sB_A(Z, D_f)$ is the specific binding energy of a cluster per nucleon, $m_{str}(D_f)$ is the structural inertia of the system, $u_{D_f} = \dot{D}_f$ is the rate of variation of the fractal dimension, and $p_{D_f} = \frac{\partial L_{str}}{\partial \dot{D}_f} = m_{str}(D_f)R_0 u_{D_f}$ is the momentum of the system corresponding to its structurization.

Analogously to the Hertz principle, the principle of dynamical harmonization can be represented as the requirement of a minimum of the functional, being the length of a world line of particles in the Finsler space-time. As a result, we obtain the following statement of the variational principle of evolution: *the evolution of a system with constraints occurs along geodesic lines in the Finsler space-time with the curvature tensor corresponding to the evolution of internal constraints of the system, which are harmonized as a response to the coherent acceleration caused by the action of mass entropic forces.*

## 4. ELECTROPHYSICAL ASPECTS OF THE INTERACTIONS OF PARTICLES AND RADIATION WITH VACUUM

First, it is pertinent to present the citation from [44]: "A reasonable staring point at the consideration of the problem of many bodies would be the question about the number of bodies for the problem to be posed. …. The persons interesting in the exact solutions can find the answer, by looking at the history. For the Newton mechanics in the 18-th century, the problem of three bodies was unsolvable. After the creation (about the year 1910) of general relativity theory and quantum electrodynamics (about the year 1930), the problems of two bodies and a single one became unsolvable as well. In the modern quantum field theory, we meet the unsolvable problem without bodies (vacuum). So that if we are interested in the exact solutions, the zero number of bodies is too much".

We note that the system of particles in a longitudinal electromagnetic field can form a noninertial reference system under a coherent acceleration, if the coefficient of domination exceeds 1. In this case, the system becomes nonequilibrium and open, in particular, for the interaction with vacuum.

The gravitational force acting between all bodies is the most known mass force. The modern theories of gravitation are based on general relativity theory developed by Einstein in 1915 [45]. The Einstein theory of gravitation is founded on the following assertions:
- The density and the pressure of a substance make the space-time curved;
- the motion of particles in a curved space-time occurs along the geodesic curves and reflects the influence of the gravitation on the dynamics of particles.

The space-time is curved in volumes of the space occupied by matter, but it becomes also curved in a vicinity of bodies due to the elasticity of the space-time. The equations for $R_{ik}$ (the tensor of space-time curvature) were obtained by Einstein firstly in the form, where the source in these equations was only $T_{ik}$ (the tensor of energy-momentum of matter). Then the equations were modified by the introduction of an additional source $\Lambda g_{ik}$ that is a cosmological term describing the antigravity:

$$R_{ik} - \frac{1}{2}g_{ik}R - \Lambda g_{ik} = \frac{8\pi G}{c^4}T_{ik}. \qquad (4.1)$$

In the middle of the 1960s, E. Gliner associated the Einstein cosmological term with vacuum, whose observed energy density $\rho_V$ is determined by the cosmological constant $\Lambda$ (see, e.g., [20-21]): $\rho_V = \frac{\Lambda c^4}{8\pi G}$. The value of $\Lambda$ is not given by theory, and it can take any value that is consistent with experiment.

In the last decades, the cosmological consequences of the introduction of $\Lambda$ were experimentally confirmed, and the following assertions are considered to be proved:
- Vacuum (dark energy) dominates in the Universe; by the energy density, vacuum exceeds all "ordinary" forms of matter taken together;
- dynamics of the cosmological expansion is guided by the antigravity;
- cosmological expansion accelerates, and the space-time becomes, in this connection, static.

In the worls by E. Gliner, the processes of accelerated expansion of matter were first connected with the antigravity of vacuum, and the creation of matter with quantum fluctuations of vacuum, which are caused by the acceleration. *Vacuum should be considered as a medium occupying all the space uniformly with good reliability from the cosmological scales down to centimeters.*

The equation of state of vacuum, i.e., the connection between the pressure $p_V$ and the energy density $\rho_V$,

$$p_V = -\rho_V, \qquad (4.2)$$

follows from the theory of quantum fields and the thermodynamical reasoning. Let us use the thermodynamical identity $dW_V = TdS - p_V dV$ and represent the total internal energy of vacuum in the form $W_V = \rho_V V$. For the adiabatic processes in the homogeneous vacuum, $dS = 0$, and $dW_V = \rho_V dV$. Hence, $p_V = -\rho_V$.

By the Friedman theory [46], the gravity is created not only by the density of a medium, but also by its pressure according to the relation $\rho_{eff} = \rho + 3p$. For vacuum, the density of its effective gravitational energy $\rho_G = \rho_V + 3p_V = -2\rho_V$ is negative for a positive density. In connection with the unique equation of state (4.2) (see [21]), vacuum possesses several important properties that distinguish this medium among all others:
1. This medium cannot serve as a reference system. If there are the reference systems moving relative one another, then vacuum with the equation of state (4.2) accompanies every reference system. Hence, the nonaccelerated motion and the rest relative to this medium cannot be distinguished.
2. The medium with the equation of state (4.2) is unvariable and eternal. Its energy is the absolute minimum of the energy contained in the space.
3. The medium with the equation of state (4.2) creates the antigravity.
4. Vacuum creates a force, but it does not undergo (as a macroscopic medium) any action of external gravitational forces or the own antigravity (because the densities of the inertial mass $\rho_i = \rho + p$ and the gravitational mass of vacuum $\rho_G = \rho_i$ are equal to zero).
5. Vacuum is a medium uniformly filling the space on all scales from cosmological to small (by the data of modern experiments, down to scales of the order of centimeters). Experimentally, some manifestations of an inhomogeneity of vacuum were observed at the creation of nonhomogeneities of the medium on scales of the order of one micron and less (Casimir effects).

By virtue of the above-presented properties, vacuum plays the key role not only for the gravity, but also for any mass force, by revealing itself only in the noninertial accelerated reference systems. Therefore, substantiated is the assumption that the most important role in the interaction with vacuum is played by the electromagnetic field (vector potential) and the fields of negentropy (information) that transfer momenta to particles through the appropriate perturbations of the probability density distribution of particles in the space and create noninertial reference systems.

*The violation of the condition of adiabaticity of the equation of state of vacuum on macroscopic scales of the order of meters or centimeters or less with the help of electromagnetic and entropic drivers will allow one to control its properties on these scales and to pose the question about its implication in energetic processes.*

Changes of the entropy and the energy density on the scale of a perturbation of vacuum appear due to the action of mass forces and, hence, changes of the impactness:

$$\Delta S = \frac{\partial S_V}{\partial q}\delta q \approx \alpha_I \frac{\partial S_V}{\partial q}\delta Q_{imp}, \quad \delta \rho_V = \left(T\alpha_I \frac{\partial S_V}{\partial q}\delta Q_{imp}\right)\rho_V. \qquad (4.3)$$

**4.1. Resonances at the interaction of longitudinal waves with vacuum**

All main properties of vacuum are qualitatively determined from the equation of state and the uncertainty relation. Namely this relation reflects the peculiarities of vacuum, since it does not allow the conjugated quantities (e.g., a momentum and a coordinate or an energy and e time interval) to have simultaneously some exactly determined independent values. In this connection, the vacuum state cannot have the zero value of energy density, though it is defined as the state with minimal energy. The fluctuations of the vacuum state energy exist always, and it is impossible to get rid of them.

In a simple one-dimensional model, the fluctuational oscillations of vacuum are a collection of ideal oscillators with all frequencies. The energy density of elastic oscillations of vacuum with any frequency $\omega$ is $W_\omega = \frac{\langle p^2\rangle/m}{2} + \frac{m\omega^2\langle x^2\rangle}{2}$, where $x$ is the coordinate, and $p = mu$ is the corresponding momentum at oscillations of an oscillator with effective mass $m$. Considering the formula for the energy, as the arithmetic mean of two terms, we obtain a chain of inequalities

$$W_\omega = \frac{\langle p^2\rangle/m}{2} + \frac{m\omega^2\langle x^2\rangle}{2} \geq \sqrt{\omega^2\langle p^2\rangle\langle x^2\rangle} = \omega\left(\sqrt{\langle p^2\rangle}\cdot\sqrt{\langle x^2\rangle}\right) \geq \frac{\hbar\omega}{2}, \qquad (4.4)$$

where we use the uncertainty relation: $\Delta x \Delta p_x \geq \frac{\hbar}{2}$ on the last stage.

It follows from (4.4) that the energy minimum for oscillations of the oscillator turns out to be $W_{\min} = \frac{\hbar\omega}{2}$. The total energy density of all oscillations $W_0$ is equal to the integral contribution of all real frequencies from zero to infinity and, naturally, is infinite. Let us introduce a large, but finite scale $L$ along a separate direction. Then the continuous set of frequencies becomes a discrete infinite sequence $\omega_n = n\frac{\pi c}{L}$, and $W_0(L) = \pi\frac{\hbar c}{2L}\sum_{n=1}^{\infty} n$.

Under the action of mass forces, the adiabiticity of vacuum can be broken, and an inhomogeneity $l_R$ can appears. It will lead obviously to resonances due to a change of boundary conditions. As a result of the appearance of resonances between an electromagnetic field and vacuum, the infinite discrete sequence $\omega_n$ is separated from the continuum of frequencies. In this case, the total energy density of vacuum is infinite as before, but it is equal now to the infinite sum over all discrete frequencies $\omega_n = n\frac{\pi c}{l_R}$ or over all wavelengths $\lambda_n = \frac{\pi c}{\omega_n}$ (with regard for the resonance conditions $\lambda_n = \frac{l_R}{n}$):

$$W_0(l_R) = \pi \frac{\hbar c}{2l_R} \sum_{n=1}^{\infty} n \ . \tag{4.5}$$

Thus, the appearance of the space-time curvature causes a change of the energy:

$$\Delta W = W_0(L) - W_0(l_R)\frac{l_R}{L} = \sum_{n=1}^{\infty}\left(\frac{\pi\hbar c}{2l_R}n\right) - \sum_{n=1}^{\infty}\left(\frac{\pi\hbar c}{2l_R}\frac{l_R^2}{L^2}n\right) \ . \tag{4.6}$$

The subsequent calculations are carried on within a regularizing procedure, which allows us to find the infinite sums with the help of the introduction of the efficient "cutting" of high harmonics and the use of the relations

$$\sum_{n=1}^{\infty} W_n \to \lim_{\lambda \to 0} \sum_{n=1}^{\infty} \exp(-\lambda W_n) W_n \quad \text{and} \quad \sum_{n=1}^{\infty} n\exp(-xn) = \frac{1}{4sh^2(x)} \underset{x \to 0}{=} \left(\frac{1}{x^2} - \frac{1}{12} + \frac{x^2}{240} + \ldots\right) \ . \tag{4.7}$$

First, we calculate difference (4.6) with a finite parameter $\lambda$. Then, by passing to the limit, we obtain the formula for the difference of energies in the one-dimensional case:

$$\Delta W \approx -\frac{\pi}{24}\frac{\hbar c}{l_R} \ . \tag{4.8}$$

In the three-dimensional case, the similar calculations were first performed by Casimir [22], who considered two plane surfaces with area $S_{surf}$, which are placed at the distance $d$ from each other, and obtained th formula

$$\frac{\Delta W}{S_{surf}} \approx -\frac{\pi^2 \hbar c}{720 d^3} \tag{4.9}$$

and, respectively,

$$F_R / S_{surf} \approx -\frac{\pi^2 \hbar c}{240 d^4} \tag{4.10}$$

for forces acting on the unit area of a plate (Casimir forces).

The longitudinal electromagnetic waves with wavelength $\lambda$ that realize the coherent acceleration $a_{cog} \approx \alpha_d \frac{u_T}{\tau_{eff}} \approx 8\alpha_d \frac{u_T}{\tau} \approx 8\alpha_d \frac{u_T}{\lambda/c} \approx 8\alpha_d \frac{cu_T}{\lambda}$ of the separated subsystem of particles induce, in correspondence with (3.13), the space-time curvature with characteristic scale $l_R$:

$$l_R \approx \frac{c^2}{\sqrt{2}a_{cog}} \approx \frac{\lambda}{\sqrt{2}\cdot 8 \cdot \alpha_{d0}(A)\beta_T}, \quad \alpha_{d0}(A) = \frac{\Delta p}{p_T} \approx \frac{e \cdot A}{p_T} \ . \tag{4.11}$$

Using the resonance conditions for wavelengths $\lambda_n = \frac{l_R}{n}$ and formula (4.9) for the space-time curvature scale, we obtain the formula for the resonance frequencies:

$$\omega_{0n} \approx \frac{1}{n}\frac{4\sqrt{2}\pi}{c}a_{cog} \approx \frac{1}{n}2^{3/2}\beta_T\alpha_d\left(\frac{2\pi}{\tau_{eff}}\right) \approx \frac{2^{3/2}\beta_T}{n}\alpha_d\omega_0 \ . \tag{4.12}$$

The coefficient of domination is proportional to the amplitude of the electromagnetic field in the medium and must take the growth of the amplitude at the resonance interaction of the field with the medium into account. At a resonance, the frequency dependence of the amplitude is as follows:

$$\alpha_d \approx \frac{\alpha_{d0}}{\sqrt{\left(1-\frac{\omega^2}{\omega_{0n}^2}\right)^2 + \left(\frac{\delta_{eff}}{\omega_{0n}}\right)^2}} \approx \frac{\alpha_{d0}}{\sqrt{\left(1-\frac{\omega^2}{\omega_{0n}^2}\right)^2 + \frac{\omega^2}{4\omega_{0n}^2}\frac{1}{Q^2}}} \ . \tag{4.13}$$

Here, $\delta_{eff}$ is the damping coefficient, $\Delta\omega$ is the width of a resonance, and $Q \approx \frac{\omega_0}{\Delta\omega}$ is the quality of a resonance. From whence and (62), we obtain the formula for frequencies

$$\omega_{0n} \approx \frac{2^{3/2}\beta_T}{n} \frac{\alpha_{d0}}{\sqrt{\left(1-\frac{\omega^2}{\omega_{0n}^2}\right)^2 + \frac{\omega^2}{4\omega_{0n}^2}\frac{1}{Q^2}}} \omega_0 \approx \frac{2^{5/2}\beta_T}{n}\alpha_{d0}Q\omega_0 \approx \frac{2^{5/2}\beta_T}{n}\alpha_{d0}Q\omega_0,$$

which yields

$$\omega_{0n} = \left(\frac{c\tau_{eff}}{\sqrt{2}\alpha_{d0}\beta_T\delta_d}\right)^{1/3}\delta_d. \tag{4.14}$$

Hence, we can conclude that, for large values of the coefficient of domination $\alpha_{d0}$, the resonance frequency shifts to the lower frequencies.

*Curvature and impedance.*

It was mentioned in the previous section that, near a growing crystal and in the region of phase transitions or in the noninertial reference system in a more general case, the refractive index (or, in other words, the impedance directly connected with this physical quantity) is changed in the space. In the works by Podosenov (see [29]), the influence of constraints in electrophysical systems on the radiotechnical (electrophysical) elements entering their composition such as capacities and inductances were analyzed in details.

As was discussed above, vacuum in a noninertial system can be represented by a countable number of ideal fluctuating oscillators which can be modeled by circuits including capacities and inductances. These oscillators depend on the space-time curvature.

For a spherically symmetric motion of a system of charged particles with constraints, the Riemann space-time metric is determined by the acceleration $a_0 = \frac{eE}{2m}$, where $E \approx -\frac{\partial A}{\partial t}$ is the intensity of a longitudinal electric field on the surface of a sphere. The metric takes the form

$$ds^2 = \exp(\nu)\left(dy^0\right)^2 - \exp(\lambda)dr^2 - r^2\left(d\theta^2 + \sin^2(\theta)d\phi^2\right), \tag{4.15}$$

where

$$\exp(\nu) = \left(1+\left(\frac{\kappa}{2}\right)^{1/2}d\right)^2, \quad \exp\left(\frac{\lambda}{2}\right) = -r^2\frac{\left(\frac{\kappa}{2}\right)^{1/2}d}{1+\left(\frac{\kappa}{2}\right)^{1/2}d}. \tag{4.16}$$

The dependences of capacities and inductances on the space-time curvature $\kappa$ were studied in [29]. In particular, the capacity $C(l_{eff},\kappa)$ with characteristic spatial size $l_{eff}$ and with characteristic surface area $S_{eff}$ was obtained as

$$C(l_{eff},\kappa) \approx C_0 \frac{\left(\frac{\kappa}{2}\right)^{1/2}l_{eff}}{1-\exp\left(-\left(\frac{\kappa}{2}\right)^{1/2}l_{eff}\right)} \approx C_0\left(1+\left(\frac{\kappa}{2}\right)^{1/2}l_{eff}\right) \approx C_0\left(1+\frac{a_{cog}}{2c^2}l_{eff}\right), \quad C_0 = \varepsilon_l\frac{S_{eff}}{4\pi l_{eff}}. \tag{4.17}$$

We now determine a change of the impedance $\Delta Z(\kappa)$ of an oscillatory circuit with the capacity $C$ and the inductance $L$ at the given frequency $\omega$ near a resonance at a change of the space-time curvature $\kappa$. We take into account that the active part of the impedance tends to zero. к нулю. Hence, we can approximately write

$$\Delta Z(\kappa) \approx \frac{1}{2}\left(\omega\frac{\partial L}{\partial \kappa} + \frac{1}{\omega C(\kappa)}\frac{\partial C}{\partial \kappa}\right)\Delta\kappa \approx \frac{1}{\omega C(\kappa)}\frac{\partial C}{\partial \kappa}\Delta\kappa \approx \frac{l_{eff}}{\omega\kappa^{1/2}}\Delta\kappa. \tag{4.18}$$

The significance of relations (4.17) even for small values of currents and voltages consists in that the devices including these electrophysical elements become basically nonlinear objects such as parametric oscillatory systems. At the certain choice of the excitation frequencies, such effects as an extension of the spectrum and the amplification of a signal can be revealed.

On the other hand, a more important circumstance can possibly consist in that a deviation of the impedance from the values determined by the capacity and the properties of a dielectric in the frame of linear electrodynamics can serve a measure of the space-time curvature.

On the basis of his theory of the time as a physical quantity possessing a density [27], Kozyrev constructed a very exact device (Kozyrev's gage) to measure the changes in the time density with the use of a bridge scheme consisting of resistors and a sensitive galvanometer. By essence, the changes of the time flow in the space-time are inseparably connected with a change of its curvature. In our experiments, we used a modified scheme of Kozyrev's gage with amplifiers instead of a sensitive galvanometer (see Fig. 4.1).

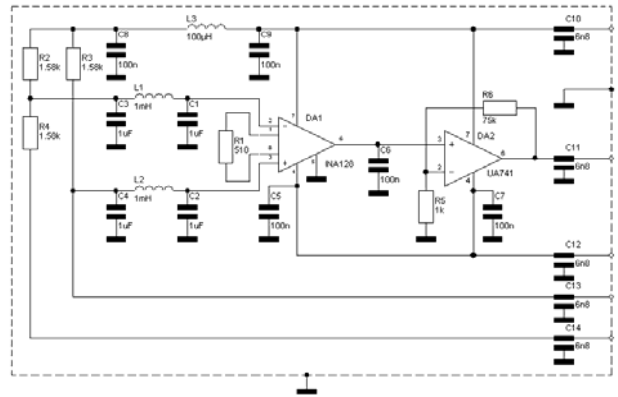

*Fig. 4.1. Kozyrev's gage with operational amplifiers instead of a galvanometer*

In the experiments carried out by Kozyrev with the use of its gage, one of the resistors of a balanced bridge served as a detector of the time flow and can be placed at various points of the region under study. A change of the impedance of this resistor was at once registered with a galvanometer. Kozyrev's gages were used as a tool in astrophysical studies by Kozyrev himself [27] and by other researchers [28].

In our studies, Kozyrev's gage was used as a sensitive meter of the space-time curvature and the appropriate resonance phenomena described above at the interaction of longitudinal electromagnetic fields with vacuum.

We have carried out the experiments on measuring the space-time curvature with the use of electromagnetic fields. As a source of longitudinal eelctric fields, we took the toroidal coil with a winding, on which a high-frequency current was flowing (see Fig. 4.2).

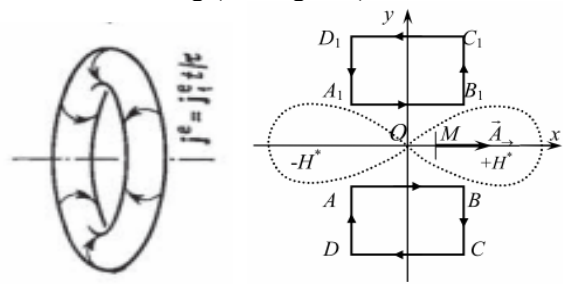

*Fig. 4.2. Toroidal coil as a source of the variable vector potential and, as a consequence, of a longitudinal electric field.*

As was expected, the maximal values of the rate of variation of the vector potential were observed on the axis of the toroidal coil.

The amplified signal from the bridge, which was proportional to a change of the impedance in the studied region where we mounted a detector, was supplied to a computer through an analog-digital unit. The changes of the impedance and, hence, the curvature were essentially different at different frequencies of a longitudinal field. The resonance frequencies were clearly distinguished.

The switching-on of a generator was naturally accompanied by an increase of the temperature in the region, where Kozyrev's gage was located. In Fig. 4.3, we show the time dependence of the measured voltage on the bridge, as well as the time dependence of the temperature.

The plots indicate clearly the complete absence of correlations of the temperature and the readings of a gage. At the time moment of the switching-on of a current in the coil, we observed a sharp change of the impedance. At the switching-off of the current, the values of signals from a bridge approach the initial values, whereas the temperature varies significantly slower.

The amplitude of variations of the impedance and, hence, the curvature demonstrated a strong dependence on the frequency of a current flowing along the emitter. The resonance behavior of the amplitude as a function of the frequency is shown in Fig. 4.5.

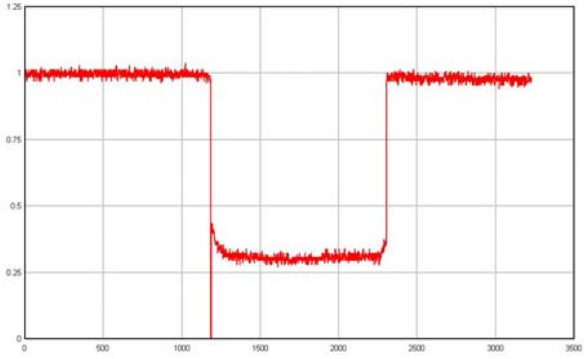
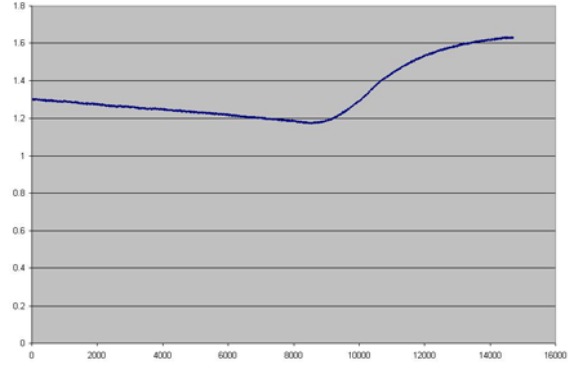

*Fig. 4.3. Temporal dependence of a signal from Kozyrev's detector. The signal is proportional to a change of the impedance and the space-time curvature.*

*Fig. 4.4. Behavior of the temperature during the measurement.*

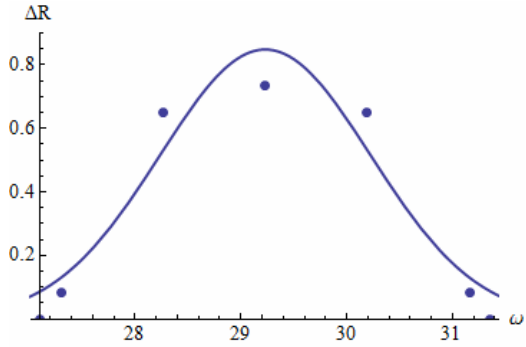

*Fig. 4.5. Resonance curve for one of the resonances in the region of frequencies of the order of 29 MHz.*

The values of the resonance frequency and the quality of resonances observed in experiments are in good agreement with (4.11) and (4.12).

**4.2. Regularized wave equations as a model of vacuum**

The physical vacuum, as a medium with specific properties, interacts with particles located in it. This interaction can accelerate particles or decelerate them. In the last case, we can say that the particles moving with acceleration undergo the action of friction due to the fluctuations of vacuum.

As was shown above (Section 2), the system of particles evolves mainly in NRS, and the efficient driver of mass forces initiating NRS is an electromagnetic driver. Below, we will study some peculiarities of the evolution of a system of charged particles at the interaction with vacuum under the action of electromagnetic fields with electric field intensity $\vec{E} = -\dfrac{\partial \vec{A}}{\partial t} - \nabla \varphi$. Only the first term in this formula is responsible for the initiation of mass forces in the system (see Section 2), since only this term represents the electric field inside a homogeneous system. We consider the action of the first and second terms on charged particles as the mass force $\vec{F}_m = -e\dfrac{\partial \vec{A}}{\partial t}$ and the Coulomb force $\vec{F}_c = -e\nabla\varphi$, respectively.

It is known [47] that the account for collisions between charged particles (electrons) in the approximation with the Landau collision integral allows one to describe the appearance of the friction force (from the side of ions), which rapidly decreases with increase in the velocity of electrons $\vec{F}_{fr} \approx -\dfrac{m_e}{4\pi} L^{e/i} \dfrac{n_i}{u^3} \vec{u}$ for velocities larger than the thermal one. In this case, it is significant that the friction force has a maximum over velocities, $\max F_{fr} \approx -0.2 \frac{e^2}{r_D^2}$, after the averaging.

For low fields, the condition of domination (2.16) does not hold. Then the charged particles obey the phenomenological equation of charge transfer or the equation of motion ($\tau_{eff}$ is the effective duration of the momentum transfer in collisions) in terms of ordinary derivatives:

$$m_e \frac{d\vec{u}}{dt} = e\vec{E}_c - \frac{1}{\tau_{eff}} (m_e \vec{u}). \quad (4.19)$$

For a quasistationary state of charged particles from the noncoherent part that are characterized by a constant velocity and satisfy the condition $e\vec{E}_c - \frac{1}{\tau_{eff}}(m_e \vec{u}) \approx 0$, Eq. (4.19) yields the Ohm law for the current density $\vec{j} = \rho_e e \vec{u}$:

$$\vec{j} = \sigma_E \vec{E}_c, \quad \sigma_E = \frac{\rho_e e^2 \tau_{eff}}{m_e}. \quad (4.20):$$

The presence of homogeneous longitudinal fields in a system of charged particles satisfying condition (2.16) corresponds to the appearance of mass forces $\vec{F}_m$. In fact, the condition of domination of the action of an electromagnetic driver (2.16) corresponds to the presence of an electromagnetic force in the system particles, which exceeds the critical friction force ($\vec{F}_m > \max \vec{F}_{fr}$). A part of charged particles (particles forming a coherent subsystem), whose share is equal to the order parameter $\eta$, are unboundedly accelerated. Another part of particles $(1-\eta)$ turns out noncoherent, is decelerated by the friction force $\approx -\frac{1}{\tau_{eff}}(m_e \vec{u})$, and obeys the equation with ordinary derivatives (4.19).

In the system, two components appear in the general case: the coherent component with density $\rho_{cog}$ and the corresponding velocity $u_{cog}$ and the noncoherent one with density $\rho$ and velocity $u$. The ratio of components determines the order parameter $\eta = \frac{\rho_{cog}}{\rho_{cog} + \rho}$, and the velocities of the components satisfy different equations of motion.

The differential equations with ordinary derivatives (Riemann derivatives), which were used for the description of the processes of transfer, are based on the fact that the translations in the space-time are characteristic ttransformations for IRS.

The satisfaction of the condition of domination (2.16) and, hence, the transition into a state with sharply decreasing resistance correspond to the appearance of a motion of the coherent part of the system as the whole with explosively increasing drift velocity (in other words, to the formation of NRS). In NRS for the systems в coherent states, the characteristic features are the presence of many scales and the self-similarity of the processes of evolution, which is reflected in a complicated (fractal, in the general case) structure of the space-time. In this case, the use of alternative definitions of the operators of differentiation, which appear due to the regularization, for the description of the dynamics of a physical situation seems to be more adequate [30].

As was shown above, the physical vacuum under conditions of the action of mass forces is characterized by a discrete set of frequencies and, hence, scales of the time. The coherence appearing in vacuum can possess the properties of similarity (fractal properties). In this case, the Jackson derivative is the most natural generalization of the notion of derivative for the description of the evolution of all quantities with the properties of similarity [31]. Let us consider the definition of this derivative, which is used, in particular, for the determination of the rate of processes. The operator of shift is replaced by the operator of scaling (with the coefficient of similarity $q_s$) passing in the limit into the ordinary derivative $D_t$:

$$D_{q_s} f(t) = \frac{f(q_s t) - f(t)}{q_s t - t}, \quad D_t f(t) = \lim_{q_s \to 1} D_{q_s} f(t). \quad (4.21)$$

The eigenfunction of the Jackson operator is the scaling generalization of the exponential function, namely $e_{q_s}^t = \sum_{k=0}^{\infty} \frac{t^n}{[k]_{q_s}!}$, which satisfies the relation $D_{q_s} e_{q_s}^t = e_{q_s}^t$. Here, the Jackson $q$-number $[n]_{q_s} = \frac{q_s^n - 1}{q_s - 1} = q_s^{n-1} + ... + 1$.

The coherence of a state of the system (scaling invariance) is revealed, naturally, in the oscillatory processes. It is easy to verify that the functions that depend on the scaling parameter $q_s$ and are defined by the relations

$$\cos_{q_s}(z) = \frac{e_{q_s}^{iz} + e_{q_s}^{-iz}}{2}; \quad \sin_{q_s}(z) = \frac{e_{q_s}^{iz} - e_{q_s}^{-iz}}{2i}, \quad (4.22)$$

satisfy the relations characteristic of ordinary trigonometric functions and are the solutions of the equation for a fractal oscillator with Jackson derivatives.

These generalized scaling functions pass into ordinary trigonometric functions as $q_s \to 1$. Respectively, the difference between the former and the latter increases with the deviation of the scaling parameter from 1.

The deviation of the parameter of similarity $q_s$ from 1 reflects a degree of openness of the system, despite the absence of an explicit dissipative term in the equation. Now, the openness of the system is characterized by the indices of differential operators of quantum analysis, rather than the parameters of dissipation. In open systems, the oscillatory processes are dissipative for the parameter of similarity $q_s < 1$ or are unstable for $q_s > 1$ (see Appendix 3).

The parameters of similarity $q_s$, nonequilibrium $q$, and damping $\delta$ are connected by the relations that can be found from the condition of maximal coincidence of phase trajectories in the quadratic metric. The result of such an optimization in the region of values of the parameter of similarity $0.7 < q_s < 1.5$ give the function (see Appendix 3):

$$q(q_s) = \begin{cases} 2.023 - 1.5608 q_s + 0.5380 q_s^2, q_s \leq 1 \\ 1.7005 - 0.9234 q_s + 0.2223 q_s^2, q_s > 1 \end{cases}. \quad (4.23)$$

Let us consider the influence of the coherence of a state on the processes of transfer of charged particles in the physical vacuum under the action of mass forces. More exactly, we will obtain a generalization of the Ohm law, which will be valid for the coherent states with the coefficient of similarity $q_s$ in homogeneous longitudinal fields $\vec{E} = -\frac{\partial \vec{A}}{\partial t}$ created by a nonstationary vector potential $A(t)$.

It is clear that, with the use of the scaling transformations, the equation for the velocity of charged particles $u$ in the case under study can be written in the form

$$D_{q_s} u = -\frac{e}{m}\left(\frac{\partial A}{\partial t}\right). \quad (4.24)$$

Acting by the integral Jackson operator $\hat{I}_{q_s} \equiv D_{q_s}^{-1}$ (see Appendix 3) on both sides of this relation, we obtain $u = \frac{e}{m} \hat{I}_{q_s}\left(-\frac{\partial A}{\partial t}\right)$. From whence, we arrive at the relation between the current density $j = e\rho_e u$ and the vector potential $A(t)$:

$$j \approx \frac{\rho_e e^2}{m \tau_{eff}} \hat{I}^v(-A(t)). \quad (4.25)$$

The proposed model of the phenomenon of transfer and oscillatory processes in fractal media on the basis of the apparatus of quantum derivatives can be a mathematical foundation for the development of new radiophysical devices using the specific properties of nonlinearity and irreversibility of the fluctuations of vacuum in NRS.

We now consider the alternative phenomenological description of the motion of the coherent part of a system of charged particles without the use of quantum operators, but with the direct application of the interaction with vacuum in NRS. As was shown above, the friction of permanently accelerating particles satisfying the condition of domination caused by collisions with other particles can be neglected. However, by virtue of the fact that these charges move as the whole and form NRS, the force of their interaction with the physical vacuum turns out to be nonzero.

The situation is similar to the motion of a body in the ideal fluid. The motion of a body with constant velocity occurs freely, and the body does not feel the presence of the medium (see the d'Alembert paradox). The motion with acceleration leads to the appearance of an associated mass and the interaction with the medium, which is proportional to the acceleration.

The motion of particles in the physical vacuum subordinates the analogous laws. The motion with acceleration leads to the interaction with vacuum and the appearance of forces $\vec{F}_{vac} = \delta_{vac} \frac{d\vec{u}}{dt}$. In this case, the equation of motion of the coherent part takes the form

$$m\frac{d\vec{u}}{dt} = -e\frac{\partial \vec{A}}{\partial t} + \vec{F}_{vac} \text{ or } (m - \delta_{vac})\frac{d\vec{u}}{dt} = -e\frac{\partial \vec{A}}{\partial t}, \qquad (4.26)$$

which yields the Ohm law for the coherent part of charged particles:

$$\vec{j} = \sigma_A \vec{A}, \; \sigma_A = \frac{\rho_e e^2}{(m - \delta_{vac})}. \qquad (4.27)$$

It is seen that the obtained Ohm law coincides with the London equation [48]. On the whole, the model of the interaction of particles with vacuum under the action of mass forces is similar to the two-fluid model of superconductivity: in the coherent state, the currents of particles arise in the absence of the difference of potentials.

The value of mass defect $\delta_{vac}$ at the interaction of particles with vacuum is determined by the explosive local expansion of the space-time with a curvature corresponding to the resonance frequencies (4.10) in metric (3.12).

## 4.3. Coherent acceleration of the reference system and criteria for the initiation of a collective synthesis

As was shown above, the basic physical quantity that initiates the processes of synthesis in an ensemble of particles in correspondence with the principle of dynamical harmonization is the coherent acceleration of this ensemble of particles.

Let us find the conditions for the acceleration of NRS that will ensure the intiation of MQO from the strongly nonequilibrium shell – i.e., the condition for the initiation of a instability leading to the reconstruction of a state of the system such as the phase transition from the stable neutral substance to a quasineutral electron-nucleus plasma. For the first time, the conditions for the appearance of a positive feedback with respect to the density in a plasmoid were found by A. Vlasov in the frame of his nonlocal kinetic theory [15]: *"The binding energy is released at a decrease of the radius of the formation and turns out to be sufficient for the support of the processes of ionization. The mechanism of the processes of ionization consists in the creation of intrinsic inhomogeneous electrostatic fields, which is a consequence of the oscillatory change of the potential of interaction of ions in the space through an intermediate system"*.

As is clear from the above, the efficient model for the description of the action of mass forces in the system of many particles is given by the Schrödinger equation and the de Broglie--Bohm representation of it in the form of a system of equations for real functions. The use of the Dirac equation for an electron in the Coulomb field of the ядра and its reduction to the Schrödinger equation *allows us to describe the mechanism and the conditions for the self-ionization of an ion on the basis of the initiation of the processes of collapse of electrons with regard for relativistic corrections* (see [5, 49 - 50]).

The conditions of stability of a dense substance (e.g., a metal) are mainly determined by properties of a degenerated electron gas. The ionization equilibrium can be changed, by varying the thermodynamical parameters such as the temperature and the density.

The increase of the degree of ionization of a substance $Z$ due to a decrease of the number of electrons shielding the nucleus causes a decrease the corresponding radius of outer electron shells.

A decrease of the size of these shells due to a change of the Coulomb repulsion leads to a decrease of the equilibrium distance between ions and, hence, to a growth of the density as compared with that in the stable state $\rho_{stabZ}$:

$$\rho_{stabZ} \approx 3.784 Z^{2.049}. \qquad (4.28)$$

The increase of the density of a substance alone can cause the ionization (figuratively, we may say that the pressure "crushes" and breaks the outer electron shells). However, this process requires very high critical densities,

$$\rho_{critionZ} \approx 5.10^2 Z^2, \qquad (4.29)$$

and the appropriate pressures. These critical densities are approximately by two orders larger than those appearing at the increase of the degree of ionization by 1. The ratio of densities $\rho_{crion}/\rho_{stab}$ as a function of the ion charge $Z$ is shown in Fig. 4.8.

Thus, the instability in the ordinary state does not arise, since the increase of the density due to the previous stage of ionization is unsufficient for the further increase of the ion charge, which ensures the stability of the substance arrounding us relative to its spontaneous collapse under equilibrium conditions.

There are the external actions on a system, at which this stability is broken. In connection with that the compression of a substance is hampered by the repulsion of like charges, all physical situations that ensure a decrease of the Coulomb repulsion due to the renormalization of the Coulomb interaction increase the equilibrium density of a substance and can induce the loss of stability.

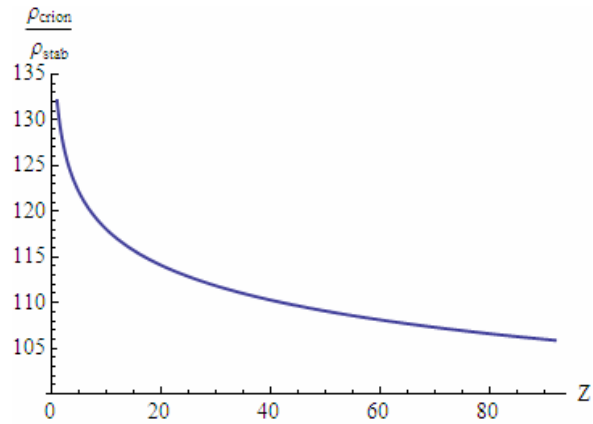

*Fig. 4.8. Ratio of the density leading to the increase of the ion charge by 1 to the density orresponding to the current ion charge*

The main contribution to the conditions of equilibrium is given by the energy of degenerate electrons. The system of particles in a thin layer (shell) becomes nonequilibrium even only due to the geometry, since two basicall different states of motion can be realized:
- perpendicularly to the thin plasma layer, the motion is bounded and, hence, has a clearly expressed discrete energetic structure;
- the motion along the layer is not bounded and has, hence, the continuous values of energy.

*In order to initiate the positive feedback leading to the collapse, it is necessary to decrease the critical density approximately by $5.10^2/3.784 = 132$ times or to delocalize the ion approximately by a factor of 5.1. Such values are available if the order parameter satisfies the following condition:*

$$\eta \geq 0.5\left(1 - \left(\frac{1}{5.1}\right)^{1/\gamma_R}\right) \approx 0.35 \text{ or } q > 1.54. \qquad (4.30)$$

The satisfaction of the conditions for existence of a positive feedback leads to the explosive process of ionization and to the appearance of an electron-nucleus plasma with the density

$$\rho_{en} \approx \frac{10}{m_p} Z_l^4 \text{ cm}^{-3}. \qquad (4.31)$$

At such a density, the mean distance between nucleons $R_{av}$ and the characteristic size of nuclei $R_{nuc}$ in a fluid are, respectively,

$$R_{av} = \left(\frac{3}{4\pi\rho_{en}}\right)^{1/3}, \quad R_{nuc} = 1.2 \cdot 10^{-13} A_l^{1/3}. \qquad (4.32)$$

Let the pressure in the environment be $p_0$. Then the collapse time of a shell can be estimated by the Rayleigh relation (the more general Zababakhin relation can be used as well):

$$t_{ex} \approx 0.9 R_{sh} \sqrt{\frac{\rho_l}{p_0}} \approx 9 \cdot 10^{-7} R_{sh} \sqrt{\frac{\rho_{lg}}{p_{0atm}}} \ . \tag{4.33}$$

The behavior of the radius tending to zero is determined by the relation

$$R_{sh} = R_{sh0}\left(1 - \frac{t}{t_{ex}}\right)^{\alpha_m}, \quad \alpha_m \approx \frac{(1+\kappa)}{2}, \quad 0 < \kappa < \infty. \tag{4.34}$$

Here, $\rho_{lg}$ is the density of the environment in gr/cm$^3$, $p_{0atm}$ is the external pressure in atm, and the shell radius in cm.

For the interaction characteristic of Maxwell molecules and hard spheres, $\kappa \approx 4/3$ and $\alpha_m \approx 7/6$. Из The analysis of solutions of the equations of dynamics of a shell yields the "scaling" relations between characteristic macroscopic scales of shells (between radii and thicknesses) and their mean densities at two time moments:

$$\frac{R_2}{R_1} \approx \left(\frac{\rho_1}{\rho_2}\right)^{3/14}, \quad \frac{d_2}{d_1} \approx \left(\frac{\rho_1}{\rho_2}\right)^{4/7} \quad \text{or} \quad \frac{\rho_1}{\rho_2} \approx \left(\frac{R_2}{R_1}\right)^{14/3}, \quad \frac{d_2}{d_1} \approx \left(\frac{R_2}{R_1}\right)^{8/3}. \tag{4.35}$$

Thus, the density at the collapse of a shell increases explosively, and the shell thickness decreases. Their behavior is described by the relations:

$$\rho_{sh} \approx \frac{\rho_0}{\left(1 - \dfrac{t}{t_{ex}}\right)^{14\alpha_m/3}}, \quad d_{sh} \approx d_{sh0}\left(1 - \frac{t}{t_{ex}}\right)^{8\alpha_m/3}. \tag{4.36}$$

The density corresponding to an electron-nucleus plasma $\rho_{en}$ is attained at the time $t_{en}$ given by the formula

$$t_{en} \approx \left(1 - \left(\frac{\rho_{en}}{\rho_0}\right)^{-\frac{3}{14\alpha_m}}\right) t_{ex}, \tag{4.37}$$

and the acceleration during the collapse increases with time.

The process of collapse of shells occurs under conditions of a dominating perturbation along the radius from the very beginning of the process (the acceleration of a coherent motion is more than the acceleration of the dissipation), whereas a decrease of the thickness acquires large accelerations only at the end of the process. At at the end of this stage, the acceleration of electrons is $a_{cog} \approx 10^{29}$ cm/sec$^2$, and the space-time curvature attains values of the order of $10^{18}$, which corresponds to the atomic scale less than $10^{-9} \div 10^{-8}$ cm.

The process of collapse in the electron-nucleus plasma, which is the fall of electrons in a Wigner--Seitz cell onto its Coulomb center (the nucleus with charge $Z$), is accompanied by the subsequent increase of the coherent acceleration up to the limitedly high values of the order of $a_{cog} \approx Z^2 \cdot 10^{29}$ cm/sec$^2$. These accelerations can already ensure the space-time curvature to be more than $10^{22}$, which corresponds to characteristic scales $\leq 10^{-11}$ cm.

The attained scales approaching the nuclear ones correspond to the high rates of change of the entropy gradient and ensure the flattening of the wave functions of all particles of the system and the formation of MQO with the scaling from the macro- down to nuclear scales.

*Thus, the self-consistent ionization of a substance due to the loss of stability caused by the action of mass forces occurs explosively and is accompanied by a change of the number of constraints in the system. The explosive change of the entropy in time and space leads to the existence of accelerations of all orders. The mass force appearing in these processes in a self-consistent way causes the explosive "flattening" of the wave functions of nuclei. If the effective size of a nucleus tends to the mean distance between nuclei, then the order parameter approaches 0.5, and MQO is formed.* The formation of MQO initiate the collective synthesis of new structures, whose efficiency depends on the dynamics of a coherent acceleration and, hence, on the space-time curvature.

## 5. CONCLUSIONS

The present work is a part of the cycle of works [1-2] devoted to the description of the theory of self-organization of the systems with varying constraints and the control over the synthesis. It is made in the frame of the development of the conception of self-organizing synthesis (see [1], [5]) on the basis of the principle of dynamical harmonization.

In the work, we have presented the geometric approach to the variational principle of dynamical harmonization, which allows one to solve the problems of self-organization and control over the directedness of the evolution of various complicated systems, basing on the single viewpoint from the very general positions of the theory of dynamical systems with varying constraints.

The comprehension of the geometric nature of physical laws was started by Clifford [51] and was developed by Hilbert, Einstein, and Wheeler [52-54].

In his mathematical works concerning the work by Riemann [55], Clifford wrote as early as 1876:
"I consider that
1. Small parts of the space are really analogous to small hills on the surface, which is plane on the average, namely: the ordinary laws of geometry do not valid there.
2. The property of curavture or deformation continuously passes from one part of the space to another one like a wave.
3. Such a change of the space curvature reflects the real phenomenon called by the motion of matter, which can be the ether or a weighty substance.
4. Only such changes obeying (possibly) the law of continuity occurs in the physical world".

Einstein analyzed the gedanken experiment with particles in the field of mass forces [53] and made conclusion that the light velocity is changed in the field of gravitational mass forces, and, hence, the space-time curvature appears. In 1919, the phenomenon predicted by Einstein was discovered experimentally during Sun's eclipse.

We note that the single property of the field of gravity, which was used in the theoretical reasoning [53], was its mass character. The analysis of the principle of dynamical harmonization and the basic positions of the conception of self-organizing synthesis [1, 5] allowed us to generalize the idea of general relativity theory of the curved space-time in the field of gravitational forces to any mass forces.

The assertion that the mass forces of any nature (satisfying the condition of domination) decrease the light velocity and curve the space-time is basic for the geometrization of the theory of evolution and control. It is worth to mention that a decrease of the light velocity in the region, where the coherent acceleration is present due to the growth of crystals, was experimentally discovered much earlier (see [25]) than a decrease of the light velocity near massive gravitating bodies.

In the frame of the theory constructed by us, we have obtained the connection between a change of the space-time curvature in quasihomogeneous electromagnetic fields and a change of the impedance (see (4.16)), which was registered in the experiment with the help of Kozyrev's gage (see [27], [28]) in the modern version.

The important circumstance for the construction of the geometrodynamics of many-scale systems with varying constraints is the following: the most important notions joining all the scales are the space-time and the entropy (or information), and the mass forces of various nature act, as usual, on the own interval of scales, but ensure the nonlocality of processes.

The comprehension of the geometric nature of nonlocality allowed Vlasov to construct a nonlocal statistical theory [14-15], which is based on the geometry of support elements – the Finsler geometry [56]. The above-presented foundations of the geometrodynamics of evolution and control for the systems with constraints belong to the series of available theories (general relativity theory and nonlocal statistical theory).

As usual, the control over dynamical systems and the optimal synthesis of new structures is realized for the system, *whose state is set by the vector in the Euclid configurational space with a given matrix of the constraint coefficients*. In this case, the control that is a vector of forces acting on the appropriate components of the system can be optimized on the basis of the solution of a variational problem with *given functional*.

In the many-scale shell model of self-organization, the situation is significantly more complicated.
1. Evolution of the systems with varying constraints occurs in the Finsler space-time. The state of the system is set by the positions of particles in the anisotropic four-dimensional Riemann space-time (base space) and by their velocities, which are tangent to the trajectories of particles at a given point and, hence, belong to the corresponding layer of the tangent bundle of the space-time. The evolution of the system, i.e., the evolution of constraints of the system, runs also in the own layer of the space-time, where the coordinates characterize the structure of the system (such coordinates are, e.g., the fractal dimension of the system or its entropy);

2. On all stages of the process of synthesis, the evolution of systems obeys always the general variational principle for the systems with varying constraints, namely, the principle of dynamical harmonization. In the geometric statement, it asserts that the system evolves always along the geodesic lines in the Finsler space-time with regard for of the constraints in the system. In this case, the optimization functional is the space-time metric defining its curvature.

3. The defining role in the efficient control over the evolution is played by the coherent acceleration (in the general case, the tensor of accelerations) in the basic Riemann space-time. The current coherent acceleration in the basic Riemann space-time determines the constraints in a system (see (3.28)) and the evolution of the system in the tangent bundle (with the fractal dimension $D_f$ as a structural coordinate in the layer) in agreement with the equations of dynamical harmonization in the Euler--Lagrange form with the corresponding Lagrange function $L_{str} = m_{str}(D_f) R_0 \frac{\dot{D}_f^2}{2} + s B_A(Z, D_f) A - U_{str}(D_f)$ (see (3.39)):

$$\frac{d}{dt}\left(\frac{\partial L_{str}}{\partial \dot{D}_f}\right) - \frac{\partial L_{str}}{\partial D_f} = 0.$$

4. Evolution of the system chabges the metric of the basic space-time (see (3.12), (3.27), (2.37), and (2.38)):

$$ds^2 = (dx^0)^2 - \sigma^2(x^0) g_{\alpha\beta}(x^1, x^2, x^3) dx^\alpha dx^\beta, \; \sigma(x^0) = exp_q\left(\frac{x^0}{c\tau_{eff}}\right), \; q(\eta) = \begin{cases} q_- = 1 - \eta, q \leq 1 \\ q_+ = \frac{1}{1-\eta}, q > 1 \end{cases}.$$

Hence, we may assert that the order parameter $\eta$ controls the space-time metric.

The control is realized by the external vector of controlling mass forces, which sets the contributions to the appropriate components of coherent accelerations of the system.
The examples of mass forces that are the most important for the self-organization (evolution) are as follows:
  1) forces of gravity and inertia;
  2) entropic forces related to the entropy production gradient;
  3) drift forces in a plasma involving the runaway of electrons in an electric field;
  4) forces arising at the polarization of vacuum, including forces of the Casimir type.

It is possible to assert that the harmonized (nonforce) control creates the space-time curvature, which is necessary for that a configuration of the system and its state will "roll down," like free ones, into the regions optimal for the realization of the process with a desired energy directedness.

By using the de Broglie--Bohm reprentation for the Schrödinger equation, we have shown the connection of nonlocality and coherence for the systems of many particles with the entropy production and mass forces. We have demonstrated that the entropic field is integral with the fields of constraints in any quantum system, in particular in MQO. Moreover, the introduction of entropic forces induces a nonlocality similar to the quantum one even in macroscopic systems. We have also analyzed the various means to create mass forces in the system and have obtained the relations for their calculation.

In a certain meaning, the space-time curvature is a hidden parameter. Since the separation of variables at the solution of the Schrödinger equation does not cause the disappearance of correlations between coordinates and momenta $r_{xp}(\kappa)$ (due to the curvature), the Schrödinger--Robertson uncertainty relation (3.25)

$$\Delta x \Delta p_x \geq \frac{\hbar}{2\sqrt{1 - r_{xp}^2(\kappa)}}$$

is valid and can be used for the control over many physical processes. For example, it would be used for the development of methods of a sharp increase of the transparency of Coulomb barriers and, hence, the probabilities of nuclear reactions [43].

The conducted studies allowed to generalize the Heisenberg uncertainty principle for energy and time in systems with variable constrains and thus with change of energy of constrains $\Delta E$, so that this ratio is directly includes the entropy change of the system (i.e., a degree of openness (see 2.26 ))

$$\Delta t \Delta E \approx \frac{\hbar}{2} \Delta S .$$

It is now quite clear that the ratio of the classical and quantum properties of the system is determined not only by the value of the Planck's constant $\hbar$, but also over the production of entropy in the system.

The developed theory of self-organization of open systems differs from the traditional nonequilibrium thermodynamics by the role of dissipation in the processes of evolution. Usually, the irreversibility of processes in a system is determined by the transition of the energy of a regular motion into the energy of a thermal random motion. In the theory with the principle of harmonization, the constraints and the structure of a system vary continuously at each hierarchical level, and the evolution is running without significant transition of energy into heat. It is obvious since one of the most important requirements to the external actions initiating the self-organizing evolution is the excess of the values of momenta of particles, which are formed by the controlling mass forces, over their thermal momenta in the system.

In the frame of the constructed geometrodynamics of the systems with varying constraints, the results obtained in works of the cycle substantiate theoretically all basic positions of the conception of self-organizing synthesis presented in [5].

Thus, the sequence of the basic processes at the evolution of the system can be presented as follows:
1. Separation of an ensemble of particles that will be evolved in the future.
2. Coherent acceleration of the ensemble and the formation of NRS as a result of the action of a dominating perturbation (creation, e.g., by electromagnetic or entropic mass forces).
3. Explosive self-consistent formation of MQO (usually of the "shell" type) when the coherent acceleration exceeds the threshold value.
4. The running of the processes of synthesis with energy directedness corresponding to the attained coherent acceleration (and, hence, to the attained space-time curvature),
5. Termination of the self-consistent process of evolution and the fixing (hardening) of products of the synthesis as a result of development of an instability on the very small scales.
6. Initiation and development of the explosive processes as a result of the release of the free energy of the synthesis of new structures.

The relations obtained in the theory of self-organization can be applied to the control over the optimal synthesis of systems with variable constraints of completely different nature and with different scale levels from nuclei and the interaction of particles with the physical vacuum to social and biological systems with complicated organization.

In many cases, a nuclear reaction is impossible because of the Coulomb repulsion of the nucleus. But the internuclear Coulomb barrier prevents only in the case when the distance between the nucleis is much greater than their de Broglie wavelength. If the de Broglie wavelength $\lambda_{DB}$ for the nuleus is longer than the distance between the core nucleis, then the MQO in quantum multipart system is being formed, and as it shown in the work, the Coulomb barriers can effectively be decreased. The possibility of increasing of the de Broglie wavelength $\lambda_{DB} = \frac{2\pi\hbar}{|\vec{p} - \vec{p}_s|}$ (up to infinity) for each particle (monomer) of the ensemble, as seen, is associated with the presence of the entropy pulse $p_s$ in them.

We emphasize once more that the base of the presented theory were main positions of the self-organizing synthesis of nuclei (see [5]), which allowed us to develop at once the means to initiate the nucleosynthesis by electron beams in a plasma diode [57]. The realization of this means allowed us to synthesize a wide spectrum of nuclei and their isomers (see [5, 2, 58-60]), and the use of electromagnetic drivers gave possibility to efficiently control the lifetime of radioactive nuclei. The experimental results concerning the electromagnetic control over the synthesis of nuclei and the rates of nuclear processes, as well as their comparison with the theory of self-organization of the systems with varying constraints, will be considered in the next article of the cycle.

The developed theory becomes rapidly a foundation for the creation of new technologies of the control over the synthesis, in particular, over the synthesis for the production of isomers-accumulators, and for the design of powerful environmentally safe "on-line" sources of nuclear energy.

**APPENDIX 1. Thesaurus of the self-organization of complex systems with varying constraints**

**Strange attractor** – set in the phase space attracting the trajectories to itself. A strange attractor has fractal structure.

**Bifurcation point –** point of branching of possible ways of the evolution of a system. In the differential formalism, the solutions of nonlinear differential equations are branched at such a point.

**Variational principles of the evolution of a system:**
- *The Hamilton principle of least action (the variational principle of the dynamics of closed systems)*

- *The Gauss principle of least compulsion (the general variational principle of the dynamics including the dynamics of the systems with constraints).* By the principle of least compulsion, the system with ideal constraints chooses the motion with the minimal "compulsion" $Z$ among all motions admitted by constraints, which start from the given position with given initial velocities. The free material point with mass $m$ under the action of a given force $F$ on it will have the acceleration equal to $F/m$. If some constraints are imposed on the point, then its acceleration under the action of the same force $F$ will be equal to a different value $w$. The deviation of the motion of the point from free motion due to the action of a constraint will depend on the difference of these accelerations $F/m - w$. The quantity $Z$ proportional to the square of this difference is called "compulsion". For a single point, $Z = \frac{1}{2}m(F/m - w)^2$

- *Hertz least-curvature principle (the variational dynamical principle, which is the closest to the Gauss principle and the most convenient for the systems with constraints).* From all trajectories admissible by constraints, the trajectory with the least curvature will be realized. This principle is also called the principle of straightest path and is closely related to the principle of least compulsion, because the quantity called the "compulsion" is proportional to the square of the curvature. For ideal constraints, both principles have the same mathematical representation.

- *Principle of minimal entropy production (Prigogine principle of evolution for dissipative systems and structures).* In 1947, I. Prigogine introduced the notions of entropy production and entropy flow, gave the so-called local formulation of the second origin of thermodynamics, and proposed the principle of local equilibrium. He showed that, in the stationary state, the entropy production rate in a thermodynamical system is minimal (Prigogine theorem), and the entropy production flor irreversible processes in an open system tends to a minimum (Prigogine criterion).

- *Principle of dynamical harmonization (the most general principle of dynamical evolution of systems with varying constraints)*

**Dissipation** – processes of energy dispersion, its transformation in less organized forms (heat) as a result of dissipative processes such as heat conduction, diffusion, etc.

**Action.** The action is the quantity $S = \int_{t_1}^{t_2} dt L(q, \dot{q}, t)$ or $S = \int_{t_1}^{t_2} dt \left( \sum_i p_i \dot{q}_i - H(q, \dot{q}, t) \right)$, where $t$ – time, $q = \{q_1, ..., q_N\}$ – complete collection of coordinates characterizing the dynamical system (its configurational space), $\dot{q} = \{\dot{q}_1, ..., \dot{q}_N\}$ – collection of velocities (derivatives of $q$ with respect to the time), $L$ –Lagrange function depending on $N$ coordinates, $N$ velocities, and, sometimes, explicitly on the time. In classical mechanics, the action coincides with the difference of kinetic and potential energies; $H$ – Hamilton function that is the total energy of the system depending on $N$ coordinates, $N$ momenta conjugated them, and, sometimes, explicitly on the time.

**Dominating perturbation** – mass force creating the coherent acceleration of particles of a system and, hence, a flow in the phase space. The value of constant flow in the phase space determines the dominating perturbation intensity for the system.

**Information.** It is intuitively assumed in the Shannon theory that information has content. Information decreases the total uncertainty and the informational entropy. The amount of information can be measured. However, Shannon warned as for the mechanical transfer of notions from its theory to other fields of science: "The search for ways of applying the theory of information to other regions of science ios not reduced to the trivial transfer of terms. This search can be realized in the long-term advancing of new hypotheses and their experimental verification."

**Coherence –** from the Latin word "cohaerentia" – internal connection, connectedness. The behavior of elements inside the system that is consistent in time and space. In physics, it is the consistent running of several oscillatory or wave processes in time and space. Coherent behavior of elements – base for the appearance of space-time structures. Coherence is continuously connected with correlations of the basic quantities in the system.

In quantum mechanics, coherent states are states with minimal disoersion (states with the probability distribution in the form of a Gauss distribution), i.e., they are states that are the closest to macroscopic states of the system.

**Coherently correlated states.** Coherently correlated states (CCS) are a complete collection of nonstationary states, in which the process of delocalization can be expanded. Обычно использующиеся The equilibrium CCS usually used for the description of the systems weakly deviating from equilibrium ones (with small accelerations and flows). To describe the processes of delocalization with limitedly large accelerations, it is necessary to use the expansions in the eigenstates of systems that are in strongly nonequilibrium states, namely nonequilibrium CCS.

**The space-time curvature** – physical effect revealing itself in a deviation of geodesic lines, i.e., in the divergence or convergence of the trajectories of freely moving bodies launched from close points of the space-time. The space-time curvature is characterized by the Riemann curvature tensor.

**Mass. Mass defect.** Mass is mainly determined by the binding energy of a system. The mass defect is a change of the mass as a result of the change of the structure of the system and its constraints.
For example:
- Mass of nucleons is determined by the binding energy of quarks;
- Mass of nuclei is determined by the binding energy of nucleons;
- Mass of a "shell" is determined by the binding energy of electrons, nucleons, and nuclei;
- Mass of atoms is determined by the binding energy of nuclei and electrons.

**Mass force** – force acting identically on all elements of a system and creating, in this case, the coherent acceleration of the system.

The example is the gravitational force acting on all particles proportionally to their masses. It is usually considered that the mass force is the reason for the appearance of a flow in the configurational space of the system.

However, in many cases where the mass force acts identically on all elements of a subsystem (separated from the whole system in some way), such a subsystem, being homogeneous in the configurational space, accelerates, i.e., a flow appears in the momentum subspace of the phase space.

The example of such a situation is given by the subsystem of electrons of a plasma in an electric field, whose intensity is more than some critical value (the runaway threshold). In this case, the plasma passes in a state with electrons running away, i.e., all electrons are coherently accelerated, and the electric field acting on the plasma plays the role of a *dominating perturbation*, which acts on the plasma and transfers the subsystem of electrons in a coherent state.

If a flow in the phase space of the system (or coherent acceleration) is not constant and is in the state with positive feedback, *the blow-up mode* arises.

**Instability by Lyapunov –** instability with respect to the initial data, which leads to the exponential divergence of earlier close trajectories.

**Lyapunov indices** – increments of the instability with respect to the initial data (instability by Lyapunov).

**Ill-posed problem** – problem, whose solution is unstable with respect to the initial data or to a perturbation of the operator.

**Nonlocality** – main characteristic of a system, being in the mode of coherent acceleration (the blow-up mode). In this case, the state of the system cannot be set by the expansion in a vicinity of the given point in infinitely small values and, hence, by the acceleration of a single order. The system is characterized by the accelerations of all orders. The property of nonlocality is characteristic of the systems in the blow-up mode, systems near a phase transition, and *MQO*.

**Blow up:**
- *Duration of the blow-up* – finite time interval, during which the process is developing with a superhigh rate.
- *Blow-up mode* – mode possessing a long-term quasistationary stage and a stage of superfast growth of the processes in open nonlinear systems. The dynamics of basic quantities in the blow-up mode is described by an explosive function $\approx \left(1 - \frac{t}{\tau}\right)^{-\nu}$ diverging at the blow-up time moment $\tau$.

**Flow in the phase space.** Usually, the flow of a physical quantity is the amount of this quantity transferred in unit time through any area in the space. For the coherently accelerating systems, whose properties are identical over the whole volume, the significant parameter is the amount of such a quantity transferred in unit time through an area in the energetic or momentum space irrespective of the coordinates. The flow in the phase space (like the coherent acceleration) is related to the degree of deviation of a state of the system from the equilibrium one corresponding to the zero flow (or, what is the same, to the zero coherent acceleration).

**Dimension of a system:**
- *Dimension of the embedding of a system* – minimal number of parameters completely describing a state of the system.
- *Fractal dimension* – fractional dimension characterizing the self-similarity and the scaling invariance of systems.

**Resonance excitation** – correspondence of the spatial and temporal structures of an external action to the internal structures of an open nonlinear system.

**Regularization. Operator of regularization.**
To obtain a stable solution of an *ill-posed problem, it is necessary* to use some special methods called the methods of regularization. It is possible to define the spaces, where the solutions of a problem become proper or, by applying *the operators of regularization* (the operators of special averaging), to change the operators defining the problem or to introduce new *observable* variables.

**Self-organization** – process of spontaneous ordering, formation, and evolution of structures in open nonlinear systems.

**Synthesis** – process of formation of new structures, i.e., the process of formation of new constraints.

**System:**
- *Open system.* System, which exchanges with the environment by energy, mass, and information.
- *Closed system.* System, which does not exchange with the environment by energy, mass, and information. Energy and information in the closed systems are conserved.

**Reference systems:**
- *Inertial reference system– reference system, in which the bodies not subjected to the action of forces move along straight lines.*
- *Noninertial reference system– reference system moving with acceleration relative to an inertial reference system*

**Structure** – set of elements of a system with a set of stable constraints between elements:
- *Dissipative structure* – stable state of an open system, which arises as a result of the dissipation of the energy continuously supplied from outside. Prigogine developed the theory of dissipative structures to explain the behavior of systems, being far from the equilibrium. In this case, the properties of the system in small regions of the space are described by locally equilibrium functions with the values of macroscopic parameters strongly different from equilibrium ones. The strong deviation from the equilibrium in dissipative structures means large spatial gradients of macroscopic parameters of a locally equilibrium system. In this case, the moving forces of the evolution are the gradients of physical quantities.
- *Nonlocal structure.* Structure, which arises as a result of the process of self-organization, i.e., the evolution of constraints in the system in its whole spatial volume, and differs from the equilibrium system even locally. The self-organization of the system is initiated by mass forces leading to the coherent acceleration (in the absence of significant gradients of macroscopic parameters inside the system). The reconstruction of constraints and their energies in nonlocal structures occurs namely due to the coherent acceleration at the dissipated energy and the gradients inside the system tending to zero.

**Fractal objects** – objects possessing the properties of self-similarity or scaling invariance.

**Phase portrait** – possible states of a system in its phase space; the set of trajectories of the system in its phase space.

**Phase space (space of states)** – multidimensional space, whose coordinates serve as parameters completely describing a state of the system.

**APPENDIX 2. Basic notation.**

$\eta$ – order parameter

$D_f$ – fractal dimension

$q$ – parameter of nonequilibrium

$q_s$ – parameter of similarity

$\alpha_d$ – parameter of domination

$Q_{imp}$ – parameter of impactness

$\tau_{eff}$ – effective duration of the operation of a driver

$\tau_{dis}$ – effective duration of the dissipation

$a_{cog}$ – coherent acceleration

$F_m$ – mass force

$F_{str}$ – mass force initiating the formation of a structure

$\sigma_S$ – entropy production

$J$ – action

$S$ – entropy

$u_T$, $p_T$ – thermal velocity and momentum

$m_{str}$ – structure inertia (structure mass)

$m$ – mass

$B$, $sB$ – binding energy and specific binding energy per nucleon

$L_{str}$ – Lagrange function at the formation of a structure

$Z_{dh}$ – dynamical harmonization functional

$\vec{A}$, $\varphi$ – vector and electrostatic potentials

$A$, $Z$ – mass and charge of a nucleus

$l_+$ – delocalization scale

$l_-$ – localization scale

$\delta$ – deformation

$\delta(q_s)$ – damping decrement or increment of the instability

$D_{q_s}$ – Jackson operator with the parameter of similarity $q_s$

$Q$ – quality of an oscillatory circuit

$\kappa$ – space-time curvature

$g_{ik}$ – space-time metric

$R_{ik}$ – Riemann curvature tensor

## APPENDIX 3. Main relations for the Jackson operators (integro-differential operators of quantum analysis)

The fractal media are characterized by the properties of the similarity of basic quantities at a variation of the space scales. Therefore, The most natural generalization of the notion of derivative is the Jackson derivative [4], in which the scaling operation (with the coefficient of similarity $q_s$) is used for the determination of the rate of a process instead of the operators of shift:

$$D_{q_s} f(x) = \frac{f(q_s x) - f(x)}{q_s x - x}. \qquad (1)$$

In the limiting case, the Jackson derivative passes to the ordinary one: $Df(x) = \lim_{q_s \to 1} D_{q_s} f(x)$.

The question arises: Which functions are the eigenfunctions of the operators of Jackson $q$-derivatives? On the basis of the development of the notion of $q$-derivatives, the so-called quantum analysis was constructed, in the frame of which the generalizations of many significant mathematical relations were found. For example, let us calculate the quantum $q$-derivative of a power function:

$$D_{q_s} x^n = \frac{(q_s x)^n - x^n}{(q_s - 1)x} = \frac{q_s^n - 1}{q_s - 1} x^{n-1} = [n]_{q_s} x^{n-1}, \qquad (2)$$

where $[\alpha]_{q_s} = \frac{q_s^\alpha - 1}{q_s - 1}$ is the Jackson $q$-number, whose limits are $\lim_{q_s \to 1}[\alpha]_{q_s} = \alpha$ and $\lim_{q_s \to \infty}[\alpha]_{q_s} = q_s^{\alpha-1}$.

It is simple to calculate the derivative of a function possessing the property of similarity. Let $f(q_s x) = q_s^\alpha f(x)$, then

$$D_{q_s} f(x) = \frac{(q_s^\alpha f(x)) - f(x)}{(q_s - 1)x} = \frac{q_s^\alpha - 1}{q_s - 1} \frac{f(x)}{x} = [\alpha]_{q_s} \frac{f(x)}{x} \qquad (3)$$

The eigenfunction of the Riemann derivative is the exponential function $e^x$, which can be expanded in a power series $e^x = \sum_{k=0}^{\infty} \frac{x^k}{k!}$, where $k!$ is a factorial. Quantum analysis uses widely the $q$-generalization of the exponential function $e_q^x$, whose power expansion contains the generalization of $k!$, which is replaced by $[k]_{q_s}!$:

$$[k]_{q_s}! = \begin{cases} 1, & k = 0 \\ [k]_{q_s} [k-1]_{q_s} \cdots [1]_{q_s}, & k \geq 1 \end{cases}. \qquad (4)$$

In other words, the power series for the $q$-exponential function takes the form

$$e_{q_s}^x = \sum_{k=0}^{\infty} \frac{x^n}{[k]_{q_s}!}. \qquad (5)$$

It is easy to see that such a definition implies that the function $e_q^x$ is the eigenfunction of the operator $D_q$:

$$D_q e_{q_s}^x = D_{q_s} \left( \sum_{k=0}^{\infty} \frac{x^k}{[k]_{q_s}!} \right) = \sum_{k=0}^{\infty} \frac{1}{[k]_{q_s}!} D_{q_s}(x^k) = \sum_{k=1}^{\infty} \frac{[k]_{q_s}}{[k]_{q_s}!} x^{k-1} = \sum_{k=1}^{\infty} \frac{1}{[k-1]_{q_s}!} x^{k-1} = e_{q_s}^x \qquad (6)$$

The quantum derivative is a linear operator. Therefore, the $q$-derivative of a linear combination of functions can be presented in terms of the derivatives of separate functions by the ordinary relation. However, the $q$-derivative of a product of functions has already some specific features.

Definition (1) yields the relations for the derivatives of a product of functions that differ from ordinary relations by the absence of symmetry. Namely, two different relations are simultaneously valid for the derivative of a product of functions:

$$D_{q_s}(f(x)g(x)) = f(q_s x)D_{q_s}g(x) + g(x)D_{q_s}f(x)$$
$$D_{q_s}(f(x)g(x)) = f(x)D_{q_s}g(x) + g(q_s x)D_{q_s}f(x). \tag{7}$$

For the functions possessing the similarity, $f(q_s x) = q_s^\alpha f(x)$ and $g(q_s x) = q_s^\beta g(x)$, we obtain

$$D_{q_s}(f(x)g(x)) = f(x)D_{q_s}g(x) + q_s^\beta g(x)D_{q_s}f(x) =$$
$$= f(x)D_{q_s}g(x) + g(x)D_{q_s}f(x) + (q_s^\beta - 1)g(x)D_{q_s}f(x). \tag{8}$$

Hence, the parameter of similarity $q_s$ of the quantum differentiation characterizes simultaneously the degree of its asymmetry.

In addition, quantum analysis considers the operators, which are inverse to the derivatives – the operators of $q$-primitives. The function $F(x)$ is called the $q$-primitive for a function $f(x)$, if $D_q F(x) = f(x)$, and is denoted by $\int f(x)d_q x$. It is easy to see that if a function $f(x)$ is set by a power series $f(x) = \sum_{k=0}^{\infty} a_k x^k$, then

$$\int f(x)d_{q_s} x = \sum_{k=0}^{\infty} \frac{a_k}{[k+1]_{q_s}} x^{k+1} + C. \tag{9}$$

Sometimes, it is convenient to use the formal definition of the Jackson integral for the $q$-primitive of a function $f(x)$:

$$\int f(x)d_{q_s} x = (1-q_s)x\sum_{k=0}^{\infty} q_s^k f(q_s^k x). \tag{10}$$

We note that the Jackson $q$-numbers $[x]_{q_s}$, which are expressed in terms of the parameter of similarity $q_s$, are closely related to the Tsallis nonextensive entropy for the states with the parameter of nonequilibrium $q$: $S_q = -\sum_i p_i^q \ln_q(p_i) = \dfrac{1-\sum_i p_i^q}{q-1}$. In the definition of entropy, we apply the generalized logarithm $\ln_q x = \dfrac{x^{q-1}-1}{q-1}$, which satisfies the relation $\ln_q(xy) = \ln_q(x) + \ln_q(y) + (1-q)\ln_q(x)\ln_q(y)$.

The main property of the generalized entropy $S_q$ consists in that it is not already the extensive function. If the system is divided into two independent subsystems A and B, then

$$S_q(A+B) = S_q(A) + S_q(B) + (1-q)S_q(A)S_q(B). \tag{11}$$

Deviations from the symmetry and the ideality in this relation are determined, like that in (8), by the deviation of the relevant parameter from 1.

The majority of equilibrium physical parameters of closed ideal systems are expressed via ordinary exponential functions coinciding with their generalized analogs for the coefficient of nonequilibrium $q \approx 1$ and the coefficient of similarity $q_s \approx 1$. The degree of deviation from the thermodynamic equilibrium and the ideality is determined by the deviation of the mentioned parameters from 1. The nonideal states of the system must manifest themselves, naturally, in the oscillatory processes, which are realized in fractal media.

**A model of oscillatory processes in fractal media on the basis of quantum analysis**

To study the peculiarities of oscillatory processes in fractal media, we consider the generalization of the trigonometric functions on the basis of $q$-exponential functions (5). In the frame of quantum analysis, the following new functions are introduced:

$$\cos_q(z) = \frac{e_q^{iz} + e_q^{-iz}}{2}; \quad \sin_q(z) = \frac{e_q^{iz} - e_q^{-iz}}{2i}. \tag{12}$$

Using relations (6) for the quantum derivative of a generalized exponential function, it is easy to obtain that the functions introduced with the help of relations (12) satisfy the relations similar to the relations for trigonometric functions:

$$D_q \cos_q(z) = -\sin_q(z), \quad D_q \sin_q(z) = \cos_q(z). \tag{13}$$

These $q$-functions pass into the ordinary trigonometric functions as $q \to 1$. However, the deviation of the former from the latter increases with the deviation of the parameter of nonextensity $q$ from 1. In Fig. 1, we present the plots for the $q$-trigonometric functions $\sin_q(t)$ for various parameters of similarity.

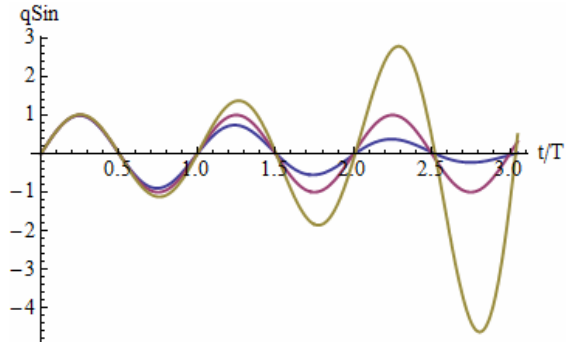

Fig. 1. Simplest oscillatory process in a fractal medium. Plots of the functions $\sin(t)$ and $\sin_q(t)$ are given for the parameters of similarity $q_s = 0.95$ and $q_s = 1.05$

As is seen from Fig. 1, the oscillatory process described by $q$-trigonometric functions has character of a dissipative process. Let us analyze this analogy in more details. Consider the simplest self-similar oscillatory process, which is described by the simple equation for a fractal oscillator with the use of the Jackson derivatives:

$$D_q(D_q f(x)) + \omega^2 f(x) = 0. \tag{14}$$

By the direct substitution, it is easy to verify that the general solution of this equation is the function $f(x) = C_1 \sin_q(\omega x) + C_2 \cos_q(\omega x)$. The case shown in Fig. 1 corresponds to the initial conditions $f(0) = 0$ and $D_q f(0) = 1$, for which $f(x) = \sin_q(\omega x)$.

In practical applications, it is convenient to approximate the Jackson $q$-functions, which are represented by infinite series, by their finite algebraic expressions. It is natural to make it with the use of nonequilibrium quasipower generalizations of the exponential function, $\exp_q(x) = (1 + (1-q)x)^{\frac{1}{1-q}}$, which allow us to write the quasipower generalizations of the trigonometric functions:

$$qCos(z) = \frac{\exp_q(iz) + \exp_q(-iz)}{2}; \quad qSin(z) = \frac{\exp_q(iz) - \exp_q(-iz)}{2i}. \tag{15}$$

We now consider the generalized exponential functions $\exp_q(-z_k)$ and $e_{q_s}^{-z_k}$ on the interval $0 \le z_k \le 4$ and define the connection between the parameter of nonequilibrium $q$ and the parameter of similarity $q_s$ from the condition of minimum for $\sum_{k=1}^{N}\left(\exp_q(-z_k) - e_{q_s}^{-z_k}\right)^2$. As a result, we obtain

$$q(q_s) = \begin{cases} 2.023 - 1.5608 q_s + 0.5380 q_s^2, & q_s \le 1 \\ 1.7005 - 0.9234 q_s + 0.2223 q_s^2, & q_s > 1 \end{cases}. \tag{16}$$

In Fig. 2 on the left, we show the self-similar oscillatory process $f(x) = \sin_q(\omega x)$ and its approximation with the generalized exponential functions $\exp_q(-z_k)$, for which the parameter $q$ is determined by relation (16). We indicate a sufficiently high accuracy of the approximation. On the right, we present the phase portrait of this self-similar oscillation.

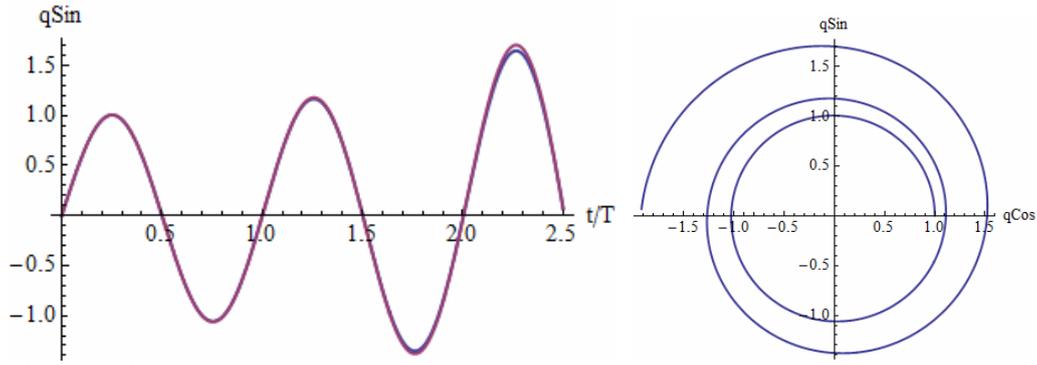

*Fig. 2. Simplest oscillatory self-similar process, its approximation with generalized exponential functions (on the left), and the corresponding phase trajectory (on the right).*

It is seen that the oscillatory process with the coefficient of similarity $q_s$ corresponds approximately to the unstable oscillatory process $g(x) = e^{-\delta x}\sin(\omega x + \Delta\varphi)$ described by the equation with ordinary derivatives for an oscillator with negative damping $\delta$:

$$D_x^1\left(D_x^1 g(x)\right) - \delta D_x^1 g(x) + \omega^2 g(x) = 0. \tag{17}$$

The parameters of similarity and damping are connected by a relation that will be determined from the condition of the maximal coincidence of the phase trajectories of a self-similar oscillation and unstable (or decaying) linear oscillations by the method of least squares:

$$\delta(q_s) = \begin{cases} 3.4931(1-q_s)^{0.6473}, & q_s < 1 \\ -10.8126(q_s-1)^{0.7969}, & q_s > 1 \end{cases} \tag{18}$$